\documentclass[review]{elsarticle}

\usepackage{hyperref}
\usepackage{amsmath}
	\usepackage{makeidx}
	\usepackage{amsfonts}
	\usepackage[ansinew]{inputenc}
	\usepackage{subfigure}
	\usepackage{epsfig}

\journal{Journal of \LaTeX\ Templates}

\begin{document}

\begin{frontmatter}

\title{Wave mixing in coupled phononic crystals via a variable stiffness mechanism}


\author{Gil-Yong Lee}
\address{Aeronautics and Astronautics, University of Washington, Seattle, Washington, 98195 USA}

\author{Christopher Chong}
\address{Department of Mathematics, Bowdoin College, Brunswick, ME 04011, USA}

\author{Panayotis Kevrekidis}
\address{Department of Mathematics and Statistics, University of Massachusetts, Amherst, Massachusetts 01003-4515, USA}

\author{Jinkyu Yang\corref{corresponding}}
\address{Aeronautics and Astronautics, University of Washington, Seattle, Washington, 98195 USA}

\cortext[corresponding]{Corresponding author}
\ead{jkyang@gmail.com}

\begin{abstract}
We investigate wave mixing effects in a phononic crystal that couples the wave dynamics of two channels -- primary and control ones -- via a variable stiffness mechanism. We demonstrate analytically and numerically that the wave transmission in the primary channel can be manipulated by the control channel's signal. We show that the application of control waves allows the selection of a specific mode through the primary channel. We also demonstrate that the mixing of two wave modes is possible whereby a modulation effect is observed. A detailed
study of the design parameters is also carried out to optimize the switching capabilities of the proposed system. Finally, we verify that the system can fulfill both switching and amplification functionalities, potentially enabling the realization of an acoustic transistor.
\end{abstract}


\end{frontmatter}


\section{Introduction}

Phononic crystals (PCs) are defined as periodic elastic lattices that can transmit or block a selected range of mechanical vibrations (i.e., phonons)~\cite{sound_atten, PC1, PC4, PC2, mitigate1, PC3}. Wave propagation in PCs has been a subject of intense study for the past several years \cite{gran1, gran2, PC5, Martin}. Particularly, the controllability and tunability of PCs has gained significant interest given the potential of  
acoustic devices that are counterparts of electrical/optical flow control devices \cite{local_reson1, switch1}. For example, researchers have shown controllable vibration mitigation and wave attenuation in PCs by tuning their local resonances \cite{mitigate1, local_reson2, local_reson3}. Such locally-resonant PCs are often called acoustic/mechanical metamaterials. 
Recent studies have also demonstrated the feasibility of wave manipulation \cite{switch1, switch2, TET1}, mixing \cite{2D_PC, HPC, 3D_wood}, and energy localization \cite{TET1, TET2,brightb,darkb} by leveraging the tunable properties of PCs. 

Many studies have focused on the investigation of tunable wave propagation in PCs in a static manner (e.g., by changing mass \cite{tune1}, stiffness \cite{HPC, tune2}, and other pre-determined mechanical properties \cite{tune3}). The more challenging problem is to manipulate wave propagation \textit{in situ} in a dynamical manner. 
Only a limited number 
of studies have been conducted to explore the dynamic control of wave modes in the realm of PCs. 
Vakakis \textit{et al.} investigated the coupling interactions of mechanical waves in nonlinear lattices via a targeted energy transfer concept~\cite{TET1, TET2}. Also, wave mixing effects have been explored in the settings of birefringent phononic crystals~\cite{Psarobas_2014}, granular wave guides~\cite{Gonella_PRL2015}, and anharmonic oscillator ladders~\cite{khomeriki}. In the majority of these systems, lattices (or particles) are interconnected with each other via \textit{fixed} linear or nonlinear springs without considering the possibility of varying the systems' stiffness. Variable stiffness may offer an enhanced degree of freedom in manipulating mechanical waves.

Traditionally, variable stiffness mechanisms have been widely used for engineering applications, such as vibration isolation systems (VIS) \cite{var_stiff1, var_stiff2}, robotics \cite{compliant1, compliant2}, and rehabilitation devices \cite{compliant3}. 
Only recently have researchers adopted variable stiffness mechanisms for the manipulations of mechanical waves in PCs. Bergamini \textit{et al.} attempted to integrate active components in PCs to adjust frequency band structures adaptively \cite{tune3}. Li \textit{et al.} leveraged helical structures to realize variable inter-lattice stiffness \cite{HPC}, which, in turn, lead to the dynamical tunability of frequency band structures. In these systems, however, the stiffness parameters of the PCs are altered by external sources and are not controlled dynamically by the application of mechanical waves.

In the present study we investigate wave dynamics in a PC that is capable of controlling one wave mode via another in a dynamical manner. This system is composed of primary and control wave channels, which are interlinked with each other via a variable stiffness mechanism. We first develop a theoretical model to characterize the wave dispersion of the primary channel in relation to the control waves. As a result, we find that the variable stiffness mechanism enables a selective transmission of two different wave modes through the primary channel, and to a further extent, a wave modulation effect. 
We show that these characteristics can be tuned by the excitation frequency, magnitude, and phase of the control wave relative to the primary one. We also verify phenomena such as mode selection, modulation, and filtering by numerical simulations of the relevant discrete particle model. Lastly, we assess the feasibility of using the proposed PC system with a variable stiffness mechanism for the realization of acoustic transistors. We confirm that the system can theoretically present both switching and amplification effects, such that the transmission efficiency in the primary channel can be easily manipulated by the control waves. 
We conclude the manuscript with a brief summary of findings and a discussion of possible future directions.

\section{Theoretical Model of Coupled Phononic Crystals With Variable Stiffness}

\begin{figure}[htbp]
	\begin{center}
		\includegraphics[width=1\textwidth]{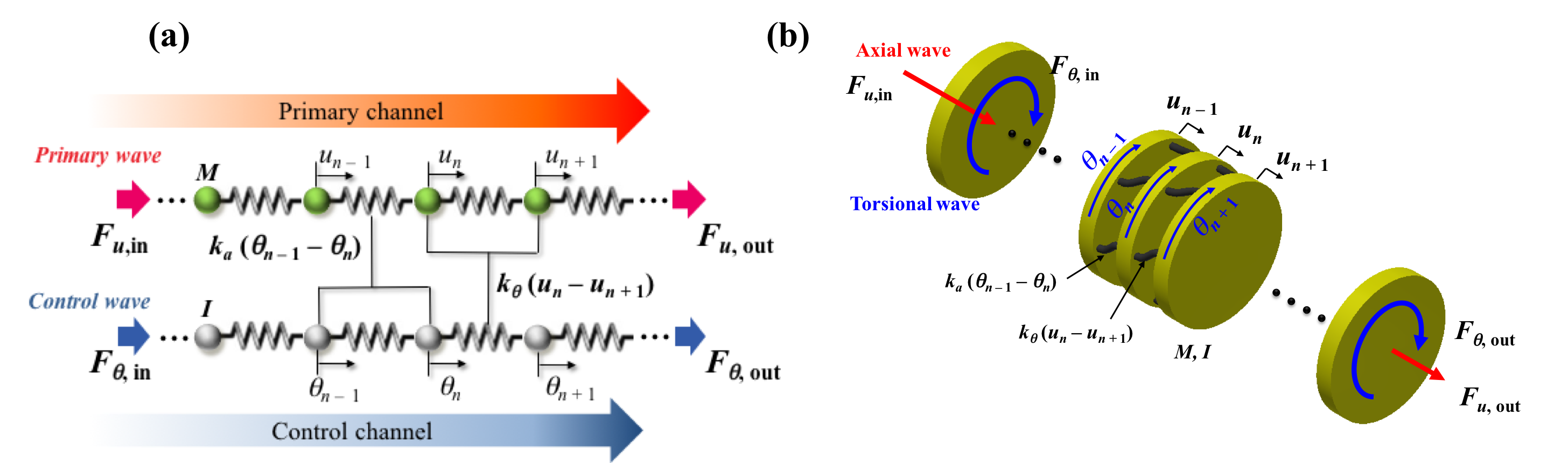}
		\caption{(a) A conceptual diagram of the coupled phononic crystal via a variable stiffness mechanism. (b) An exemplary configuration of variable stiffness system for axial and torsional wave coupling.}
		\label{fig_scheme}
	\end{center}
\end{figure}

A conceptual illustration of a coupled PC via a variable stiffness mechanism is shown in Fig.~\ref{fig_scheme}. The PC exhibits two wave transmission channels: a primary and a control route, which are represented herein by two displacement parameters $u_n$ and $\theta_n$, respectively, for the $n$-th particle (see Fig. \ref{fig_scheme}a). Note that in this system, the inter-lattice stiffness in one channel is affected by the displacement fields of the other channel. Figure \ref{fig_scheme}b shows a prototypical system to be investigated in this study, where there exists a variable stiffness mechanism between the axial and torsional directions.
Given a generalized interaction of particles under a power law, we can express inter-particle forces as $F_u \sim \delta^p$ and $F_{\theta} \sim \alpha^s$, where $F_u$ and $F_{\theta}$ are the axial and torsional force, and $\delta$ and $\alpha$ are the axial and torsional displacements, respectively. $p$ and $s$ are the power-law factors that determine the force relationship (e.g., $p = s = $1 for a linear relationship). Based on this, the equations of motion of this system can be expressed as follows:
\begin{eqnarray}
	\label{eq:eom}
		M\ddot{u}_{n} =& k_{a, (\theta_{n-1}-\theta_{n})}[\delta_0+u_{n-1}-u_{n}]^p  -k_{a, (\theta_{n}-\theta_{n-1})}[\delta_0+u_{n}-u_{n+1}]^p, \nonumber\\
		I\ddot{\theta}_{n} =& k_{\theta, (u_{n-1}-u_{n})}[\alpha_0+\theta_{n-1}-\theta_{n}]^s  -k_{\theta, (u_{n}-u_{n-1})}[\alpha_0+\theta_{n}-\theta_{n+1}]^s.
\end{eqnarray}
\noindent Here, $M$ and $I$ represent the inertia of each particle in axial and tangential directions, and 
$\delta_{0}$ and $\alpha_{0}$ denote the axial and torsional displacements under static equilibrium, respectively. Note that the two equations above are coupled since the axial stiffness $k_{a, (\theta_{n-1}-\theta_{n})}$ is affected by torsional displacements $\theta_{n-1}-\theta_{n}$, while the torsional stiffness $k_{\theta, (u_{n-1}-u_{n})}$ is affected by axial displacements $u_{n-1}-u_{n}$. Therefore, upon the dynamic perturbations applied to the control channel, the stiffness along the primary direction $k_{a, (\theta_{n-1}-\theta_{n})}$ will vary, which in turn affects the propagation of the primary waves. While this work focuses on the theoretical and numerical investigations of such coupled systems, this variable stiffness system can be realized in various settings. These examples include compliant mechanisms \cite{compliant1, compliant2, compliant3}, cylindrical Hertzian contacts \cite{HPC, 3D_wood}, and Kresling origami \cite{origami}.

In this study, we assume that the stiffness parameters 
in the primary and control channels are governed by a power-law relationship, similar to the form of aforementioned force-displacement relationship. Mathematically, this can be expressed as $k_{a, (\theta_{n-1}-\theta_{n})} = K_a (\alpha_0 + \theta_{n-1}-\theta_{n})^r$ and $k_{\theta, (u_{n-1}-u_{n})} = K_\theta (\delta_0 + u_{n-1}-u_{n})^q$, where $K_a$ and $K_\theta$ are axial and torsional constants, and $r$ and $q$ are nonlinear factors, respectively. 
This power-law approximation is plausible, since dynamic expressions of variable stiffness mechanisms can be fit with a power-law in a local domain of the PC's vibrational motions. For example, one may consider a 
chain of buckled beam elements that connect masses and support coupled axial and lateral (tangential) motions~\cite{Timoshenko, Timoshenko_book}. 
Similarly, if we consider a chain of PC composed of spherical or cylindrical particles under the Hertz-Mindlin contact law, the contact stiffness becomes a function of the relative angular displacements between neighboring cells defined by the elliptic integrals~\cite{Hertzian2, Landau}. We provide the details of these two prototypical systems in Appendix. 

Under the power-law approximation, Eq.\eqref{eq:eom} becomes,
\begin{eqnarray}
	\label{eq:power}
		M\ddot{u}_{n} =& K_a[\delta_0+u_{n-1}-u_{n}]^p[\alpha_0+\theta_{n-1}-\theta_{n}]^r - \nonumber\\
		& K_a[\delta_0+u_{n}-u_{n+1}]^p[\alpha_0+\theta_{n}-\theta_{n+1}]^r, \nonumber\\
		I\ddot{\theta}_{n} =& K_\theta[\delta_0+u_{n-1}-u_{n}]^q[\alpha_0+\theta_{n-1}-\theta_{n}]^s - \nonumber\\
		& K_\theta[\delta_0+u_{n}-u_{n+1}]^q[\alpha_0+\theta_{n}-\theta_{n+1}]^s.
\end{eqnarray}
While we solve this nonlinear equation directly in the numerical simulations that will follow, 
for our theoretical analysis, 
we further simplify Eq.~\eqref{eq:power}. 
For this, we postulate that small perturbations are applied to the system (i.e, $|u_{n-1}-u_{n}|\ll\delta_{0}$ and $|\theta_{n-1}-\theta_{n}|\ll\alpha_{0}$) and expand and linearize the power terms in \eqref{eq:power}. 
As a result, we obtain the following linearized equations of motion:
\begin{eqnarray}
	\label{eq:lin}
		\ddot{u}_{n} =& \frac{K_a}{M}p\delta_{0}^{p-1}\alpha_{0}^{r}[u_{n-1}-2u_{n}+u_{n+1}] +\frac{K_a}{M}r\delta_{n}^{p}\alpha_{0}^{r-1}[\theta_{n-1}-2\theta_{n}+\theta_{n+1}], \nonumber\\
		\ddot{\theta}_{n} =& \frac{K_\theta}{I}q\delta_{0}^{q-1}\alpha_{0}^{s}[u_{n-1}-2u_{n}+u_{n+1}] +\frac{K_\theta}{I}s\delta_{0}^{q}\alpha_{0}^{s-1}[\theta_{n-1}-2\theta_{n}+\theta_{n+1}].
\end{eqnarray}

For further simplification, we introduce (i) non-dimensionalized displacements $\hat{u}_{n}=\frac{u_{n}}{\delta_{0}}$, $\hat{\theta}_{n}=\frac{\theta_{n}}{\alpha_{0}}$; (ii) re-scaled mass and stiffness $\frac{I}{M}=\mu$, $\frac{K_\theta}{K_a}=\sigma$; (iii) natural frequencies $\omega_a^2=\frac{K_a}{M}p\delta_{0}^{p-1}\alpha_{0}^{r}$; and (iv) additional parameters $\rho=\frac{s}{p}\frac{\sigma}{\mu}\delta_0^{q-p+1}\alpha_0^{s-r-1}$, $\epsilon_u=\frac{r}{p}$, $\epsilon_\theta=\frac{q}{s}\rho$. Consequently, we obtain the following equations of motion in the non-dimensionalized form:
\begin{eqnarray}
	\label{eq:nondim}
		\ddot{\hat{u}}_{n} =& \omega_a^2[\hat{u}_{n-1}-2\hat{u}_{n}+\hat{u}_{n+1}] +\epsilon_u\omega_a^2[\hat{\theta}_{n-1}-2\hat{\theta}_{n}+\hat{\theta}_{n+1}], \nonumber\\
		\ddot{\hat{\theta}}_{n} =& \epsilon_{\theta}\omega_a^2[\hat{u}_{n-1}-2\hat{u}_{n}+\hat{u}_{n+1}] +\rho\omega_a^2[\hat{\theta}_{n-1}-2\hat{\theta}_{n}+\hat{\theta}_{n+1}].
\end{eqnarray}
\noindent This equation set indeed indicates the coupled mechanisms of the torsional and axial motions.
We should note in passing here that a related system of two-field couplings has been proposed recently as an example of an all-phononic digital transistor~\cite{khomeriki}. While the above work presented an amplifying effect due to the nonlinearity based band-gap transmission mechanism, our studies focus on the \textit{controllability} aspect of the system enabled even at the linear level through the \textit{crosstalking} between different wave modes.
In the following sections, we will systematically investigate wave mode mixing effects in this coupled system.

\section{Analytical and Numerical Results of Wave Mixing Effects}

\subsection{Observation of wave mixing mechanisms}

\begin{figure}[htbp]
	\begin{center}
		\includegraphics[width=0.7\textwidth]{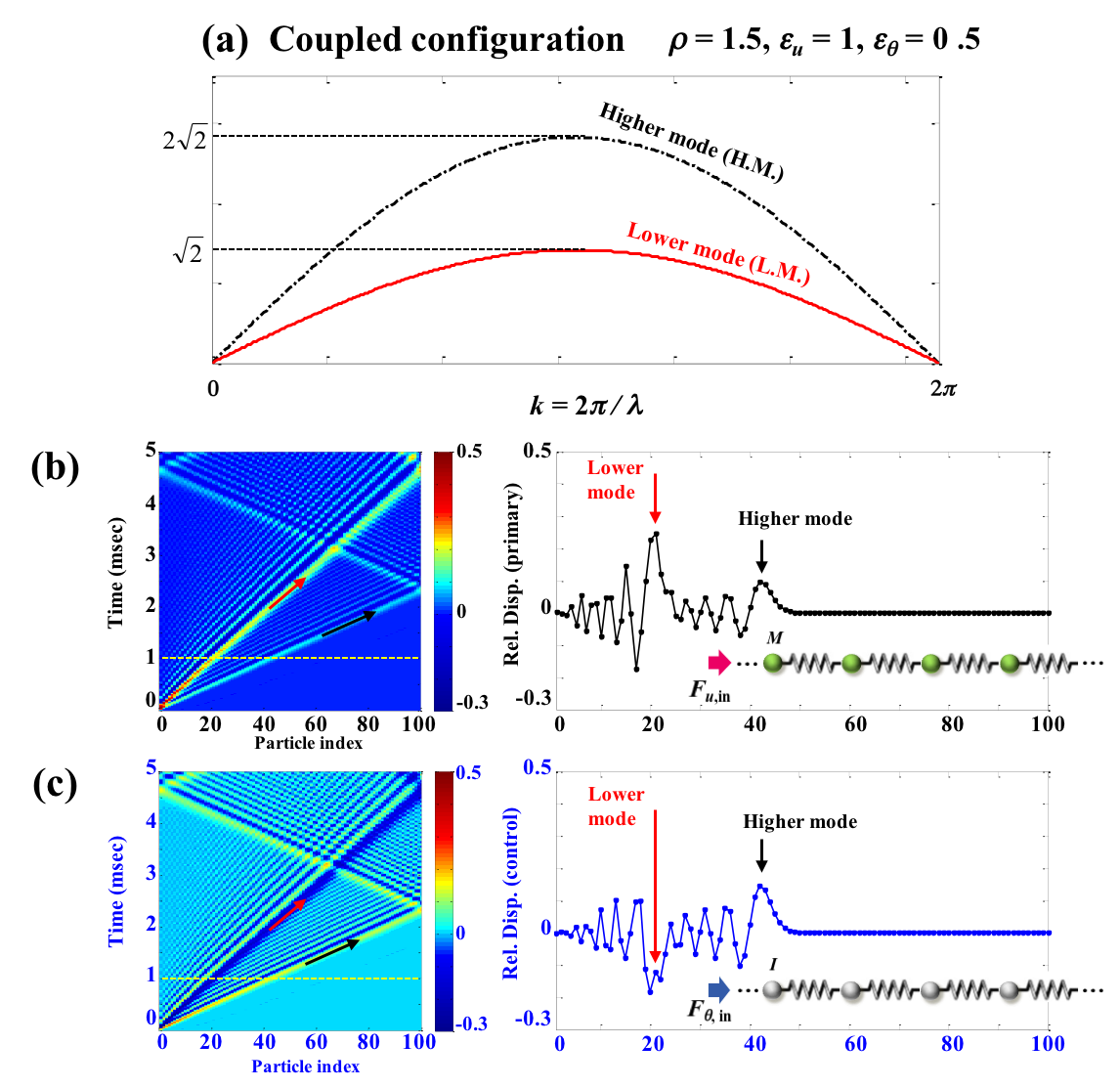}
		\caption{The analytical dispersion curves of the coupled PC ($\rho=1.5$, $\epsilon_u=1$ and $\epsilon_{\theta}=0.5$). The left panels of (b-c) represent the surface map showing the relative position of each particle as color intensity with respect to space (horizontal axis) and time (vertical axis) up on the step excitation input.
The group velocities calculated directly from the slope on the analytic dispersions in (a) were denoted as red (for primary channel) and black (for control channel) arrows. The right panels of (b-c) show the spatial wave profiles at $t = 1$ ms. The coupled PC shows that the incident waves propagate with two distinct group velocities (denoted as red and black arrows on the surface map) in the primary as well as control channels.}
		\label{fig_analytical_disp}
	\end{center}
\end{figure}

To analyze the coupling of the axial and torsional wave modes, we consider the plane wave ansatz in the form of the Fourier mode. Mathematically, we can express $\hat{u}_n=\widetilde{U}e^{j(\omega t+kn)}$ and $\hat{\theta}_n=\widetilde{\Theta}e^{j(\omega t+kn)}$, where $j=\sqrt{-1}$, $\widetilde{U}$ and $\widetilde{\Theta}$ are complex variables denoting the amplitude of axial and torsional waves, and $\omega$ and $k$ representing the angular frequency and wavevector, respectively. By substituting these expressions into Eq. \eqref{eq:nondim}, we obtain the following equations:
\begin{eqnarray}
	\label{eq:dispersion}
		{\Omega(k)}^2\widetilde{U} =& 2[\widetilde{U}+\epsilon_u\widetilde{\Theta}][1-\cos(k)], \nonumber\\
		{\Omega(k)}^2\widetilde{\Theta} =& 2[\epsilon_{\theta}\widetilde{U}+\rho\widetilde{\Theta}][1-\cos(k)].
\end{eqnarray}
\noindent We introduce the non-dimensionalized angular frequency ${\Omega(k)}^2=\frac{{\omega(k)}^2}{\omega_a^2}$ and seek nontrivial solutions of the above equations. As a result, we obtain the following dispersion relationship:
\begin{equation}
	\label{eq:dispersion_LH}
		{{\Omega(k)}^2}_{L, H}=[(1+\rho)\pm\sqrt{(1-\rho)^2+4\epsilon_{u}\epsilon_{\theta}}][1-\cos(k)].
\end{equation}
\noindent This expression represents two dispersion wave modes, denoted by subscripts $L$ and $H$ for the lower and higher branches, respectively. 
These results are similar to the ones obtained in~\cite{khomeriki}, cf.
their Eq.~(3).
Note that if $\epsilon_u=\epsilon_{\theta}=0$, this dispersion relationship is reduced to a decoupled configuration between primary and control channels, where the dynamics of the system is described by two independent dispersion relations. 

As we introduce non-zero $\epsilon_u$ and $\epsilon_{\theta}$, the landscape of the dispersion relationship starts to be altered. Particularly, the 
upper cut-off frequencies of the higher and lower dispersion branches become functions of the degree of coupling represented by $\epsilon_u$ and $\epsilon_{\theta}$:
\begin{eqnarray}
	\label{eq:cutoff} 
	{\Omega}_{C,L}=\sqrt{2(1+\rho)-2\sqrt{(1-\rho)^2+4\epsilon_{u}\epsilon_{\theta}}}, \nonumber\\
	{\Omega}_{C,H}=\sqrt{2(1+\rho)+2\sqrt{(1-\rho)^2+4\epsilon_{u}\epsilon_{\theta}}}.
\end{eqnarray}
\noindent As an example, we show in Fig. \ref{fig_analytical_disp}a the shifted locations of such cutoff frequencies in the case that $\rho=1.5, \epsilon_u=1$, and $\epsilon_{\theta}=0.5$. 

The two wave modes in this study are fundamentally different from the classical notion of acoustic and optical modes that can yield a frequency bandgap; our system generates two wave modes that pass through the origin of the frequency-wavenumber domain (see Fig. \ref{fig_analytical_disp}a). This implies that the higher branch in our system encompasses the lower one, and is thereby capable of generating mixed-mode waves under excitations. 

To study mixed-mode waves, we conduct numerical simulations of the coupled system and observe propagation of waves in the spatial domain. We specifically consider a coupled PC with 100 particles in each channel and solve Eq.~\eqref{eq:power} directly using a fourth-order Runge-Kutta time integration scheme \cite{TFT1}. 
Figures \ref{fig_analytical_disp}b-c show the spatio-temporal map (left panel) of propagating waves and their spatial profile at $t = 1$ ms (right panel) in terms of the relative displacements (i.e., $\hat{u}_n - \hat{u}_{n-1}$ in Fig. \ref{fig_analytical_disp}b and $\hat{\theta}_n - \hat{\theta}_{n-1}$ in Fig. \ref{fig_analytical_disp}c). Upon the application of a simultaneous step force inputs $F_{u, in}$ and $F_{\theta, in}$ to the system ($F_{\theta, in}=0.5F_{u, in}$ is used in this simulation), 
we find two wave modes propagating in each channel. The group velocities calculated directly from the analytical dispersion curves in Fig.~\ref{fig_analytical_disp}a (i.e., the slopes at the first wave number $k=\frac{2\pi}{N}$, where $N=100$) are in agreement with the speed of the wave packets denoted by red (for higher mode) and black (for lower mode) arrows in Figs. \ref{fig_analytical_disp}b and \ref{fig_analytical_disp}c. As one can infer from Fig.~\ref{fig_analytical_disp}a, the faster mode corresponds to the higher mode, while the slower, trailing mode represent the lower mode (see the arrows in Fig. \ref{fig_analytical_disp}b-c). As shown in the surface maps plotted in the left panels of Fig. \ref{fig_analytical_disp}b-c, we can clearly distinguish the two simultaneous group velocities in both primary- and control- channels. Therefore, by the simple onset of the coupling coefficients (i.e., $\epsilon_u$ and $\epsilon_{\theta}$), we find that the wave dispersion in the primary channel starts to be affected by the control signal, and vice versa.

\subsection{Manipulation of primary wave via control wave}

Now that we verified the effect of wave mixing, the next question is whether we can manipulate the transmission properties of the primary waves via the control waves; notice that which one is the primary and which is the control field in our setting is a matter of choice (that can be modified at will; hence it is by convention that we make the choice utilized herein). 
Specifically, we are interested in selectively choosing a specific wave mode in the primary channel by the application of control waves. To investigate this, we revisit Eq.~\eqref{eq:dispersion} and rewrite it in terms of the relative amplitude of the control input with respect to the primary input:
\begin{eqnarray}
	\label{eq:dispersion_rel}
		{\Omega(k)}^2=2[1+\epsilon_u\frac{\widetilde{\Theta}}{\widetilde{U}}][1-\cos(k)], \nonumber\\
		{\Omega(k)}^2=2[\epsilon_{\theta}\frac{\widetilde{U}}{\widetilde{\Theta}}+\rho][1-\cos(k)].
\end{eqnarray}
\noindent 
We are now interested in calculating $\frac{\widetilde{\Theta}}{\widetilde{U}}$ 
to assess the \textit{controllability} of the control signal on the primary one.

By making the two equations in Eq.~\eqref{eq:dispersion_rel} equal, we obtain two solutions of $\frac{\widetilde{\Theta}}{\widetilde{U}}$ as:
\begin{equation}
	\label{eq:control}
		{\frac{\widetilde{\Theta}}{\widetilde{U}}}_{L,H}=\frac{(\rho-1)\pm\sqrt{(1-\rho)^2+4\epsilon_u\epsilon_{\theta}}}{2\epsilon_u}
\end{equation}
Since $\widetilde{U}$ and $\widetilde{\Theta}$ are the amplitude of the primary (axial) and control (torsional) waves based on our convention, they are directly related to the primary and control force inputs, $F_{u,in}$ and $F_{\theta,in}$. 
That is, in our linear setting, if we excite the system with the excitation amplitude ratio $\frac{F_{\theta,in}}{F_{u,in}}$ identical to ${\frac{\widetilde{\Theta}}{\widetilde{U}}}_{L}$, we will trigger the onset of the lower-mode wave propagation, i.e., $\Omega(k)_L$. On the other hand, when we choose $\frac{F_{\theta,in}}{F_{u,in}}$ to be equal to ${\frac{\widetilde{\Theta}}{\widetilde{U}}}_{H}$, the wave propagation through the primary PC follows the higher mode, $\Omega(k)_H$. 


\begin{figure}[htbp]
	\begin{center}
		\includegraphics[width=1\textwidth]{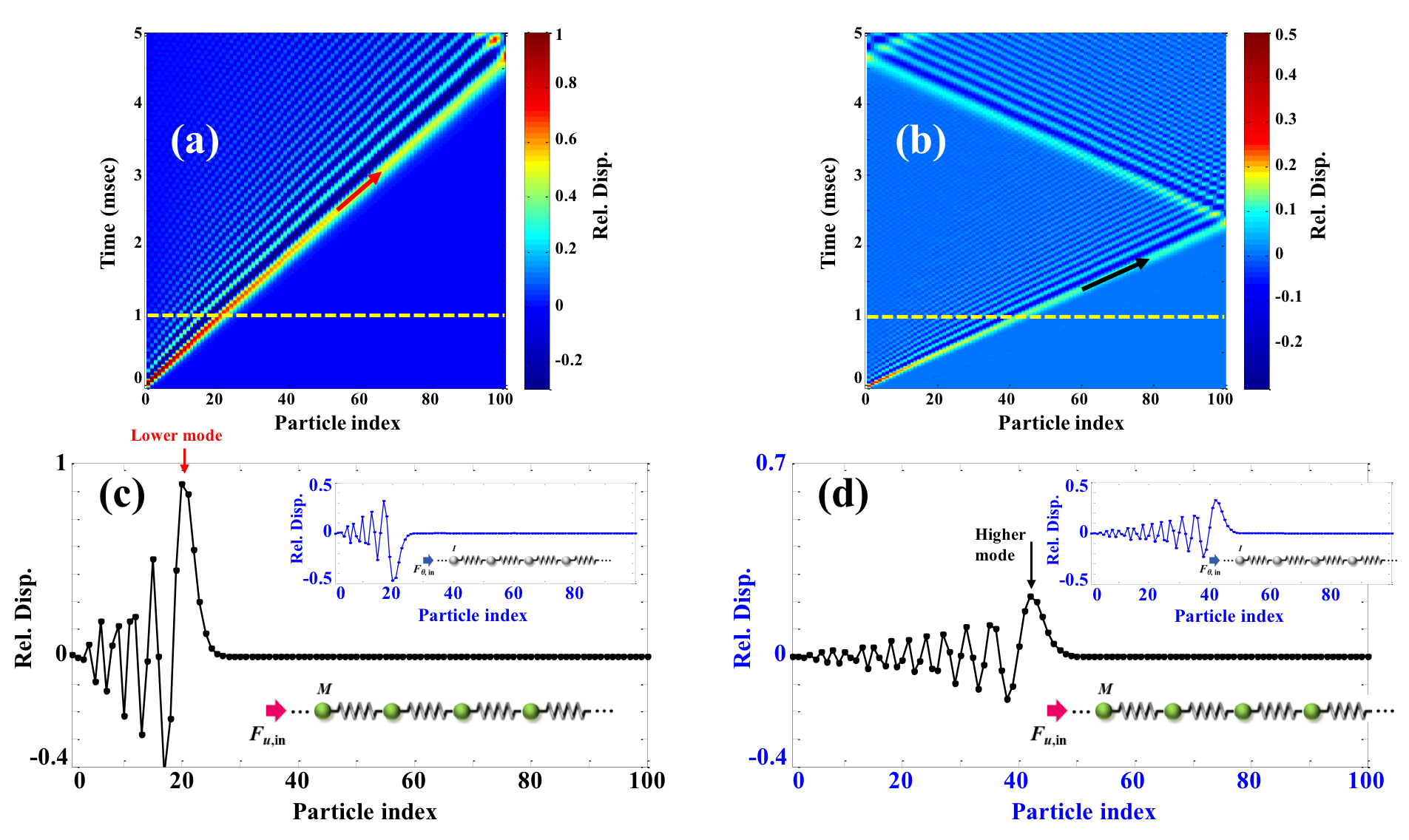}
		\caption{The evolution of wave profiles in the space and time domain (color intensity represents the relative displacements) in the primary PC, showing that the propagating wave can be controlled to support (a) the lower mode (slower group velocity) by taking the control input $\frac{F_{\theta,in}}{F_{u,in}}={\frac{\widetilde{\Theta}}{\widetilde{U}}}_{L}$ and (b) the higher mode (faster group velocity) by taking the control input $\frac{F_{\theta,in}}{F_{u,in}}={\frac{\widetilde{\Theta}}{\widetilde{U}}}_{H}$. (c-d) Spatial wave profiles at $t=1$ ms. Insets in (c-d): the spatial wave profiles of the control waves. Note that in the inset of (c), the wave front in the control PC has a 
phase difference $\pi$ with respect to that of the primary PC.}
		\label{fig_step}
	\end{center}
\end{figure}

To demonstrate this, we again solve \eqref{eq:power} directly by means of
 numerical simulations, considering the case of $\rho=1.5$, $\epsilon_u=1$, and $\epsilon_{\theta}=0.5$, which was shown to exhibit a coupling mechanism between the primary and control channels in the previous section (Figs. \ref{fig_analytical_disp}b and \ref{fig_analytical_disp}c). According to Eq.~\eqref{eq:control}, these system parameters lead to $({\frac{\widetilde{\Theta}}{\widetilde{U}}})_{L}=-0.5$ and $({\frac{\widetilde{\Theta}}{\widetilde{U}}})_{H}=1$. Note that $({\frac{\widetilde{\Theta}}{\widetilde{U}}})_{L}$ should be negative, which means that the control input $F_{\theta,in}$ should have a $\pi$-phase shift
 relative to the primary input $F_{u,in}$. In Fig. \ref{fig_step}, we plot the spatio-temporal maps (top panel) and spatial profiles (at $t = 1$ ms, bottom panel) of incident waves propagating through the primary PC as we choose the control input to be one of these two characteristic solutions. As predicted, we observe the propagation of the lower mode only, when we impose $\frac{F_{\theta,in}}{F_{u,in}} = -0.5$ (Figs. \ref{fig_step}a and \ref{fig_step}c). If we apply $\frac{F_{\theta,in}}{F_{u,in}} = 1$, we find the primary channel propagates the higher mode of the dispersive waves only (Figs. \ref{fig_step}b and \ref{fig_step}d). In both panels of Figs. \ref{fig_step}c and \ref{fig_step}d, the insets show the spatial profile of the control signals. From this simulation, it is evident that the wave mixing effect in this coupled channel enables us to select a specific mode of wave propagation. Also, note the difference of the propagating waves' speed between the two cases (depicted red and black arrows on Figs. \ref{fig_step}a and \ref{fig_step}b). 

\subsection{Switching and modulation effects}

Now we investigate the feasibility of modulating and switching output mechanical waves using the PC with a variable stiffness mechanism. For this, we first examine the frequency responses of the system by exciting each channel with a chirp signal with a broadband frequency spectrum. 
We consider three cases of the control- to primary-wave ratios: (i) $\frac{F_{\theta,in}}{F_{u,in}} = -0.5$ for the lower mode transmission; (ii) $\frac{F_{\theta,in}}{F_{u,in}} = 0.5$ for the transmission of both lower and higher modes; and (iii) $\frac{F_{\theta,in}}{F_{u,in}} = 1$ for the higher mode transmission.  

\begin{figure}[htbp]
	\begin{center}
		\includegraphics[width=1\textwidth]{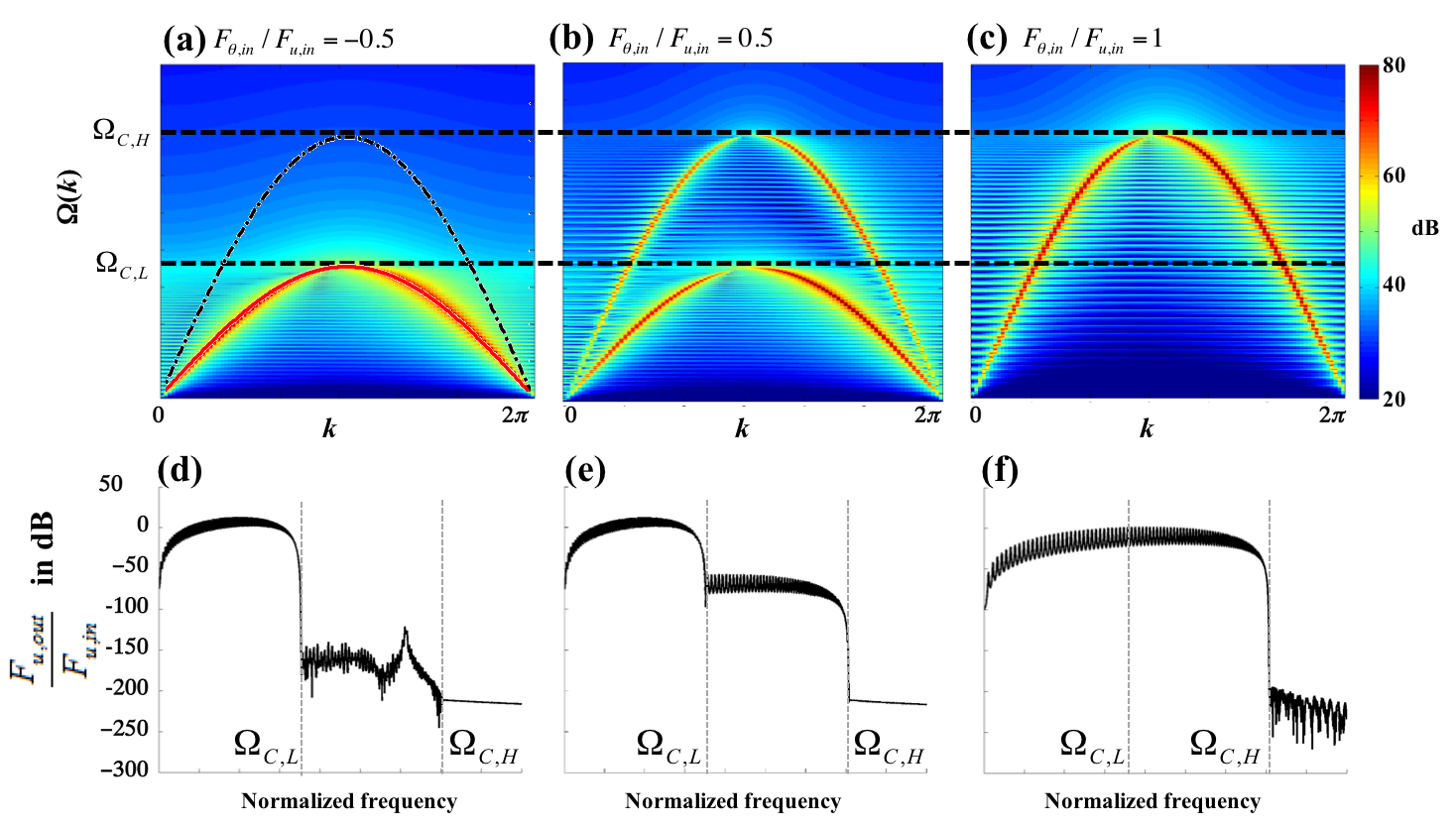}
		\caption{Power spectral density (PSD) of particles' velocities in the primary channel under three different control inputs: (a) $\frac{F_{\theta,in}}{F_{u,in}}=-0.5$, (b) $\frac{F_{\theta,in}}{F_{u,in}}=0.5$, and (c) $\frac{F_{\theta,in}}{F_{u,in}}=1$. The PSD profiles of the transmitted force as a function of frequencies are plotted in (d-f) for the corresponding three excitation conditions. The cutoff frequencies of the higher and lower modes ($\Omega_{C,L}$ and $\Omega_{C,H}$) are indicated by the dashed lines.}
		\label{fig_PSD}
	\end{center}
\end{figure}

Figure \ref{fig_PSD}a-c shows the power spectral density (PSD) of the mechanical waves transmitted through the primary channel for the three cases. We obtain these plots by conducting a two-dimensional fast Fourier transform (2D-FFT) of the particles' velocities in the time and space domains. In the first case when we excite the system with $\frac{F_{\theta,in}}{F_{u,in}} = -0.5$, we observe the dominant presence of the lower mode (Fig. \ref{fig_PSD}a). This is consistent with the result of the system's mode selection capability discussed in the previous section. Similarly, we can also characterize the frequency response of the system by calculating the force ratio of the transmitted wave relative to the input wave ($\frac{F_{u,out}}{F_{u,in}}$, see Fig. \ref{fig_scheme} for these parameters). As a result, we find the lower pass band ($0<\Omega<\Omega_{C,L}$) exhibits much higher PSD compared to that of the higher pass band ($\Omega_{C,L}<\Omega<\Omega_{C,H}$) (see Fig. \ref{fig_PSD}d). This again confirms the system's capability of selective wave transmission. The peak in the upper frequency band is generated due to the intrinsic nonlinearity of the system, since the numerical simulations are based on the nonlinear equation (Eq.~\eqref{eq:power}).


If we gradually increase the control input, we start to observe the onset of the higher mode. For example, we see the existence of the two modes clearly when $\frac{F_{\theta,in}}{F_{u,in}} = 0.5$ (Fig. \ref{fig_PSD}b). The corresponding PSD of the transmitted force also shows the presence of both modes as shown in Fig. \ref{fig_PSD}e. As we increase the control wave further to the magnitude and phase identical to that of the primary wave (i.e., $\frac{F_{\theta,in}}{F_{u,in}} = 1$), we witness that the higher mode takes over the lower mode as shown in Fig. \ref{fig_PSD}c. In terms of the relative efficiency of the transmitted wave, we find a single low pass band that covers the frequency range from 0 to $\Omega_{C, H}$ (Fig.~\ref{fig_PSD}f). Note that the dispersion relationship obtained from the 2D-FFT agrees well with the analytical dispersion curves in the previous section (compare Fig. \ref{fig_analytical_disp}b with Figs. \ref{fig_PSD}a-c presented here). During the transition between the lower- to the higher-mode wave propagation, we witness that the transmission efficiency in the frequency range of $\Omega_{C,L}<\Omega<\Omega_{C,H}$ increases gradually. 
This translates into the \textit{switching} capability of the system, which controls the transmission efficiency of waves in the primary channel by the application of the control waves.

\begin{figure}[htbp]
	\begin{center}
		\includegraphics[width=1\textwidth]{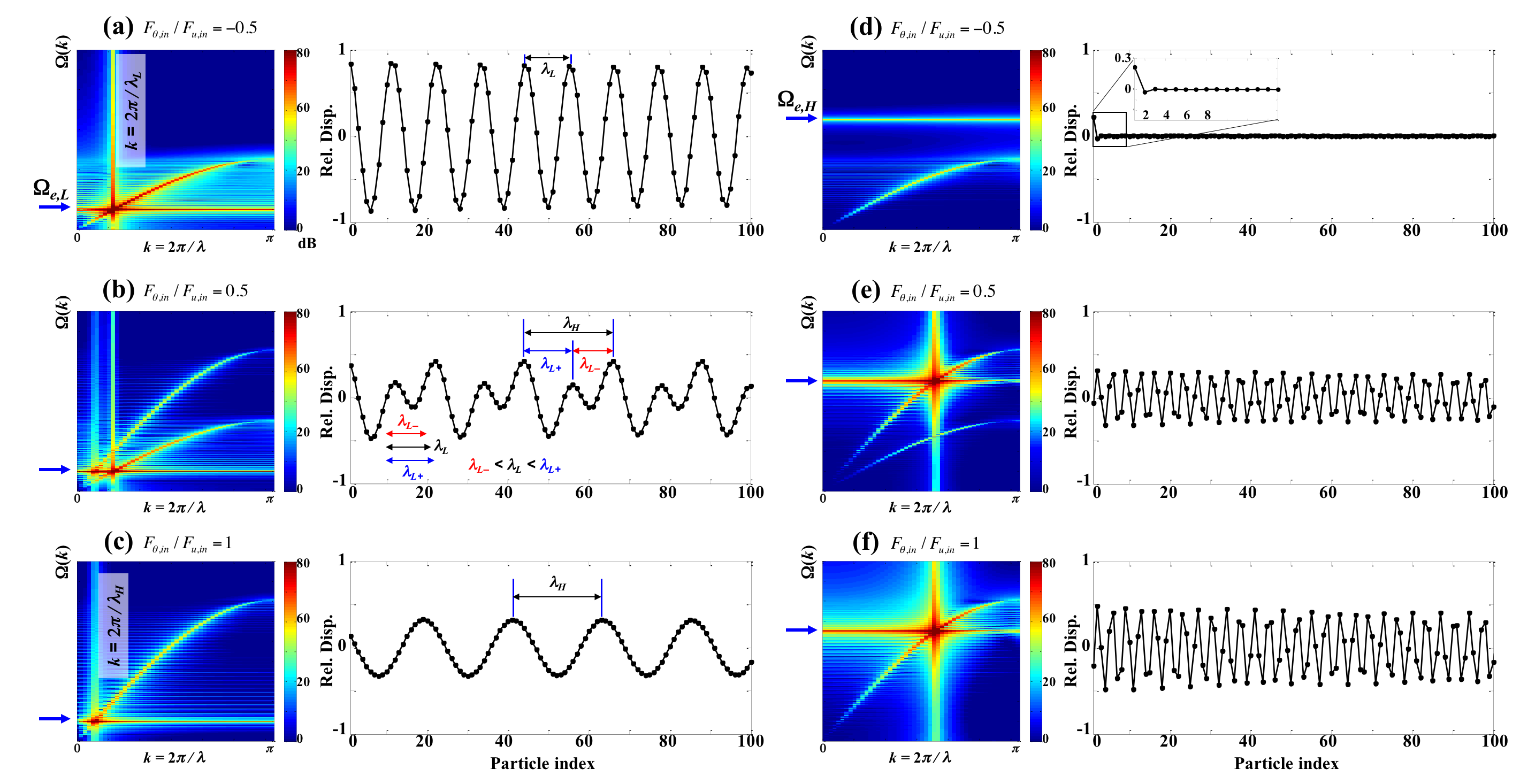}
		\caption{The 2D fast Fourier Transform (FFT) of the particle's velocities in the primary channel excited harmonically at frequencies (a-c) $\Omega_{e,L}$ and (d-f) $\Omega_{e,H}$. The wave numbers corresponding to the excitation frequencies, $\Omega_{e,L}$ and $\Omega_{e,H}$, are also denoted as $k=\frac{2\pi}{\lambda_L}$ and $k=\frac{2\pi}{\lambda_H}$ on the 2D FFT plots. The right panels of (a-f) represent the spatial wave profiles at $t=40$ ms. The excitation amplitudes of the channels are (a, d) $\frac{F_{\theta,in}}{F_{u,in}}=-0.5$, (b, e) $\frac{F_{\theta,in}}{F_{u,in}}=0.5$, and (c, f) $\frac{F_{\theta,in}}{F_{u,in}}=1$. }
		\label{fig_harmonic}
	\end{center}
\end{figure}

As presented in Fig. \ref{fig_PSD}, there are two frequency domains of interest in the presented PC system with the variable stiffness mechanism, which are $0<\Omega<\Omega_{C,L}$ and $\Omega_{C,L}<\Omega<\Omega_{C,H}$. To see how the coupled PC responds in these two frequency domains, we conduct another 
set of numerical simulations using harmonic excitations applied to the system. Specifically, we excite the system with a single harmonic function at two different frequencies: $\Omega_{e,L}$ and $\Omega_{e,H}$. These two frequencies belong to the frequency domains, $0<\Omega_{e,L}<\Omega_{C,L}$ and $\Omega_{C,L}<\Omega_{e,H}<\Omega_{C,H}$, targeting to capture the harmonic response of the system in these two 
distinct bands (See Fig. \ref{fig_harmonic}). 



Figures \ref{fig_harmonic}a-\ref{fig_harmonic}c show the harmonic responses of the system at $\Omega_{e,L}$ under the conditions of $\frac{F_{\theta,in}}{F_{u,in}}=-0.5,$ 0.5, and 1, respectively. The left panel shows 
the frequency-wavenumber response of the system, while the right panel shows captured images of particles' displacements in space. When $\frac{F_{\theta,in}}{F_{u,in}}=-0.5$, we observe an oscillatory pattern of the particles in the space domain, whose wavelength corresponds to the lower branch of the dispersion relationship (denoted by $\lambda_L$ in Fig. \ref{fig_harmonic}a). We observe a similar oscillatory wave in space, if we excite the system at $\Omega_{e,L}$ and $\frac{F_{\theta,in}}{F_{u,in}}=1$ (Fig. \ref{fig_harmonic}c). This corresponds to the dispersive wave in the higher mode, as indicated by $k=\frac{2\pi}{\lambda_H}$ in the surface map of Fig. \ref{fig_harmonic}c. In the transition regime between the two cases, however, we find an interesting phenomenon. As shown in Fig. \ref{fig_harmonic}b when $\frac{F_{\theta,in}}{F_{u,in}}=0.5$, the wave propagation through the primary channel shows a \textit{modulation} effect between the 
higher and lower modes. As shown in the spatial wave profile in the right panel of Fig. \ref{fig_harmonic}b, this modulation effect generates signals composed of two wavelengths, which are denoted as $\lambda_{L-}$ and $\lambda_{L+}$. These wavelengths satisfy $\lambda_{L-}<\lambda_{L}<\lambda_{L+}$ (compare the length of the arrows for $\lambda_{L-}$, $\lambda_{L}$, and $\lambda_{L+}$ shown in right panel of Fig. \ref{fig_harmonic}b). This is due to the mixing of the higher and lower wave modes, which is realized through this variable stiffness mechanism between the primary and control channels. This is the one of intriguing features of the coupled PC that can be exploited for engineering applications. As we expect from Eq. \eqref{eq:dispersion_rel}, the wavelength of the modulated signal depends on the control input, $\frac{F_{\theta,in}}{F_{u,in}}$, and further discussion on this can be found in Appendix B.


We also test the dynamics of the system under the excitation at $\Omega_{e,H}$ as shown in Figs. \ref{fig_harmonic}d-\ref{fig_harmonic}f. Since there exists only one possible wave mode in this frequency domain, no wave modulation effect is observed. However, by taking the control input of $\frac{F_{\theta,in}}{F_{u,in}}=-0.5$, the wave propagation through the primary channel is significantly attenuated in the frequency domain (Fig. \ref{fig_harmonic}d). The inset of the right panel in Fig.~\ref{fig_harmonic}d shows drastic attenuation of the incident wave in the form of evanescent waves. While we demonstrated the wave attenuation in a single frequency, we see that such attenuation can be achieved in broad band frequencies in the range of $\Omega_{C,L}<\Omega_{e,H}<\Omega_{C,H}$. This can distinguish our system with other active vibration control methods, which typically target single or narrow-band noise and vibration \cite{mitigate1, tune1}. As we increase the control input further to $\frac{F_{\theta,in}}{F_{u,in}}=0.5$ and 1, we observe the transmission of waves, corresponding to the higher mode of dispersive waves (Figs. \ref{fig_harmonic}e and f). Note that a similar effect of selective wave filtering has been discussed recently in a weakly nonlinear phononic system~\cite{khomeriki}. 



\section{Design Parameters of the Coupled Phononic Crystal}

\begin{figure}[htbp]
	\begin{center}
		\includegraphics[width=1\textwidth]{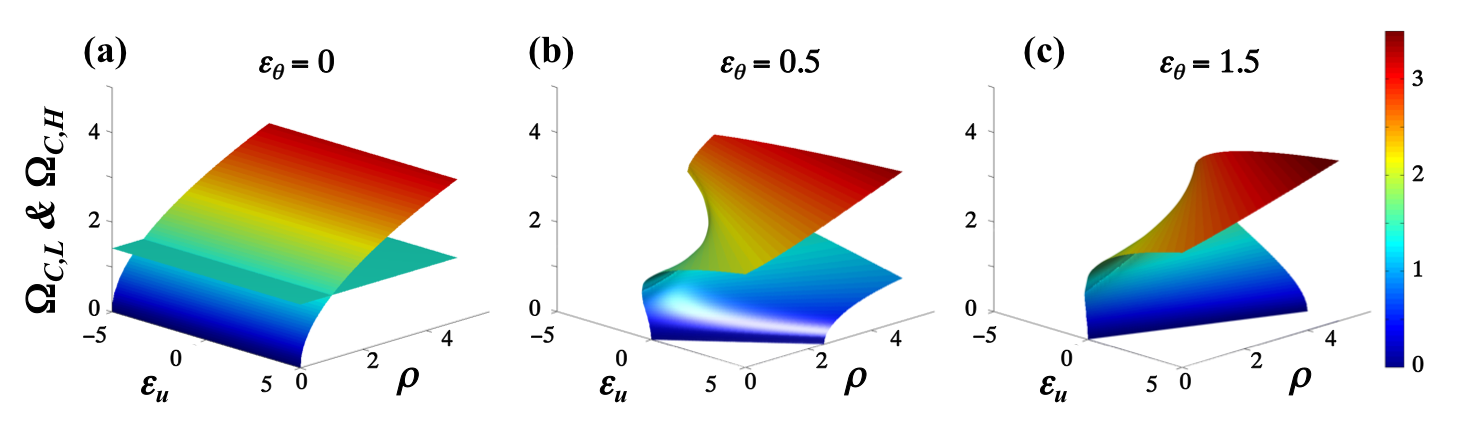}
		\caption{Surface maps of the two cut-off frequencies, ${\Omega}_{C,L}$ and ${\Omega}_{C,H}$, as a function of $\epsilon_u$ and $\rho$. We consider three different conditions: (a) $\epsilon_{\theta}$ = 0, (b) $\epsilon_{\theta}$ = 0.5, and (c) $\epsilon_{\theta}$ = 1.5. }
		\label{fig_cutoff_surf}
	\end{center}
\end{figure}

Now we investigate the evolution of the two dispersion modes under various conditions of coupling between primary (axial) and control (torsional) waves. For this, we examine the effect of coupling parameters $\epsilon_u$, $\epsilon_{\theta}$, and $\rho$ on the cutoff frequencies of the higher and lower dispersion branches. Since $\epsilon_u$ and $\epsilon_{\theta}$ play an identical role mathematically (see Eq.~\eqref{eq:cutoff}), we assess the response of cut-off frequencies as a function of $\epsilon_u$ and $\rho$ only (Fig. \ref{fig_cutoff_surf}). The three surface maps in Figs. 6a-c represent the cases when $\epsilon_{\theta}$ = 0, 0.5, and 1.5.

When $\epsilon_{\theta}=0$ (Fig. \ref{fig_cutoff_surf}a), the two dispersion curves appear with ${\Omega(k)}^2=2[1-\cos(k)]$ and ${\Omega(k)}^2=2\rho[1-\cos(k)]$ according to Eq.~\eqref{eq:dispersion_LH}. Here, we find that the two cut-off frequencies of these dispersive wave modes depend only on $\rho$, without showing any dependence on $\epsilon_{u}$ (Fig. \ref{fig_cutoff_surf}a). It should be noted that while the cut-off frequencies are not affected by $\epsilon_u$, the transmission efficiency in the primary channel are affected by the control signal. This is evident by the fact that $\epsilon_u$ is placed in the denominator of the efficiency expression in Eq.~\eqref{eq:control}. In contrast, we find that the control signal is not affected by the primary wave mode when $\epsilon_{\theta} = 0$ (see the equation of motion in Eq.~\eqref{eq:nondim}). 
As we increase $\epsilon_{\theta}$ from zero, the landscape of dispersive relationship starts to show its dependence on $\epsilon_u$, skewing the surface map of the two cut-off frequencies, ${\Omega}_{C,L}$ and ${\Omega}_{C,H}$ as shown in Fig. \ref{fig_cutoff_surf}b-c. Interestingly, we also observe some regions where a certain combination of $\rho$ and $\epsilon_u$ yields no or only a single cutoff frequency. These regions are defined by two boundaries, $\epsilon_u<-\frac{1}{4\epsilon_{\theta}}(1-\rho)^2$ and $\epsilon_u>\frac{\rho}{\epsilon_{\theta}}$, where both or one of the two cut-off frequencies in Eq.~\eqref{eq:cutoff} becomes complex. This implies that depending on the choice of system parameters, we may have (i) existence of two distinctive cutoff frequencies, where both higher and lower modes are supported (i.e., $\frac{\rho}{\epsilon_{\theta}}>\epsilon_u>-\frac{1}{4\epsilon_{\theta}}(1-\rho)^2$); (ii) existence of a single cutoff frequency, where only the higher mode can be supported ($\epsilon_u \geq \frac{\rho}{\epsilon_{\theta}}$ or $\epsilon_u=-\frac{1}{4\epsilon_{\theta}}(1-\rho)^2$); and (iii) non-existence of cutoff frequency, where the primary wave is not allowed to transmit ($\epsilon_u < -\frac{1}{4\epsilon_{\theta}}(1-\rho)^2$). A case 
example is shown in Appendix. Given this intrinsic tunability of frequency responses, we move to the next section for the feasibility study of using this system for acoustic devices.

\section{Acoustic Transistor via the Coupled Phononic Crystal}


In this section, we investigate the wave transmission property and the corresponding wave mixing effect of the proposed system. We compare the characteristic of the system's response with that of the conventional transistor, to assess its feasibility as an acoustic transistor (AT). As discussed in Section 4 and shown in Fig. \ref{fig_PSD}, the coupled PC with a variable stiffness mechanism is capable of controlling transmission efficiency of the primary channel's signals 
 in the frequency domain between the higher and lower cut-off frequencies (i.e., $\Omega_{C,L}<\Omega<\Omega_{C,H}$). When we excite this system with a frequency in this domain, the transmission efficiency through the primary channel can be altered in relation to the control signal applied to the system. 

\begin{figure}[htbp]
	\begin{center}
		\includegraphics[width=1\textwidth]{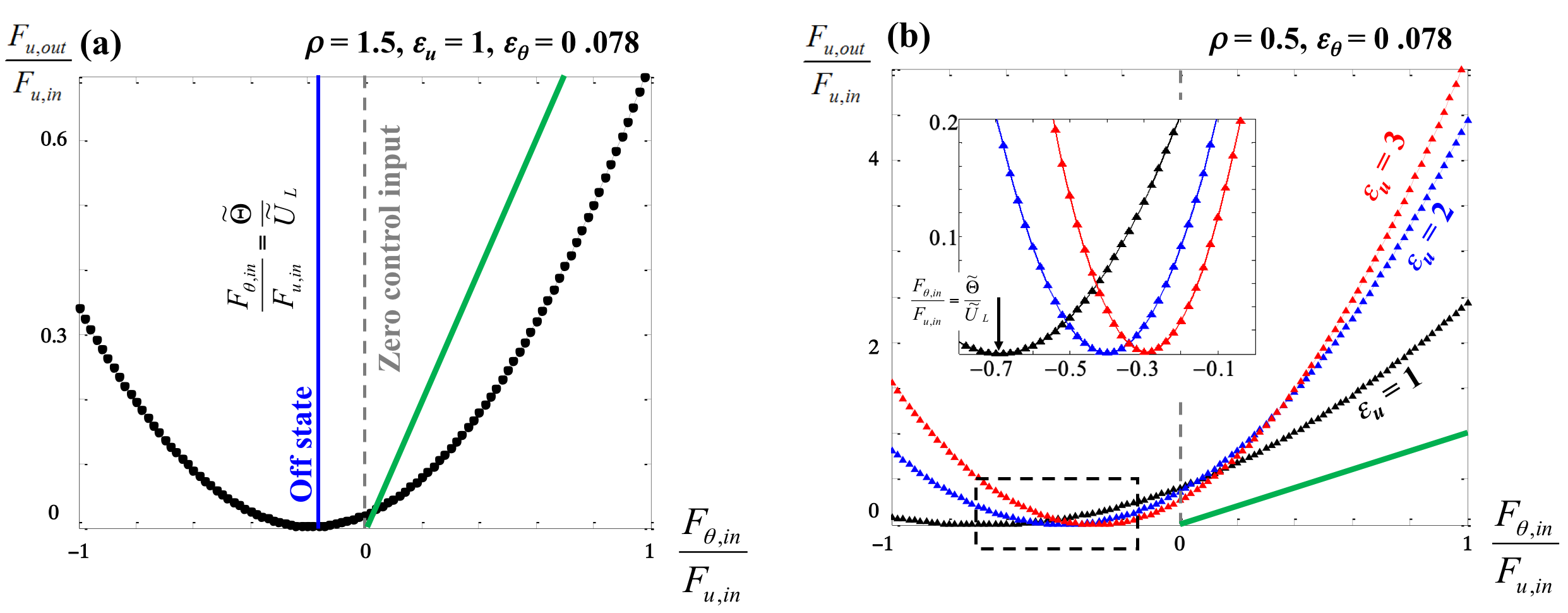}
		\caption{(a) The characteristic curve of the proposed PC system, showing the relation between the control input $\frac{F_{\theta,in}}{F_{u,in}}$ and wave transmission $\frac{F_{u,out}}{F_{u,in}}$. The system is excited with a frequency that resides in $\Omega_{C,L}<\Omega<\Omega_{C,H}$, and the parameters used are $\rho=1.5$, $\epsilon_u=$ 1,$\epsilon_{\theta}=0.078$. The solid vertical line corresponds to the off state, in which the signal in the primary channel is not transmitted. The dashed vertical line corresponds to zero control input. Transmission in this case is still possible.
The red $45^\circ$ line indicates that the control input amplitude and transmission amplitude are identical. Thus, transmission amplitude above this line corresponds to amplification. 
(b) Change of system's characteristic curves as we vary $\epsilon_u=1, 2, $ and 3. The inset shows the magnified view of the curves near the zero transmission efficiency.}
		\label{fig_AT}
	\end{center}
\end{figure}

Figure \ref{fig_AT}a shows an example of the AT effect when we use $\rho=1.5$, $\epsilon_u=1$, and $\epsilon_{\theta}=0.078$. The \textit{x}-axis denotes the control force relative to the primary wave's force amplitude, while the \textit{y}-axis represents the transmission efficiency in the primary channel, also relative to the primary wave's amplitude. To obtain this curve, the normalized amplitude of output signal $\frac{F_{u,out}}{F_{u,in}}$ is calculated with varying the control input from $\frac{F_{\theta,in}}{F_{u,in}}=-1$ to 1 with 100 steps. In Fig. \ref{fig_AT}a, we find that the transmission can be blocked by imposing the control input $\frac{F_{\theta,in}}{F_{u,in}}={\frac{\widetilde{\Theta}}{\widetilde{U}}}_{L}$ (see the vertical solid line). This corresponds to \textit{off} status, and this phenomenon is consistent with what we found in previous sections. As we change the control input from this \textit{off} state, the transmission efficiency of the primary channel gradually increases. This corresponds to the \textit{on} state of the AT. 

Interestingly, we note that at zero control input, we still have a portion of mechanical waves transmitted through the system (vertical dashed line in Fig. \ref{fig_AT}a). In the narrow region between this zero control input and the off states (i.e., the region between the two vertical lines), the increase of control signal towards the negative direction (i.e., $\pi$ phase difference between primary and control waves) reduces the efficiency of the primary channel's output. This is in contrast to the increasing trend of the transmission efficiency described in the previous paragraph. Farther past this zone towards the negative \textit{x}-axis, the increase of the control wave's magnitude results in the enhancement of the wave transmission in the primary channel. 

The plot in Fig. \ref{fig_AT}a is similar to the characteristic curve of conventional bipolar junction transistor (BJT) \cite{BJT}. While the current $i_{CE}$ from the collector ($C$) to emitter ($E$) is controlled by the electric potential $v_{BE}$ (voltage difference between base and emitter) in the electrical transistor, the proposed AT system deals with the control $\frac{F_{\theta,in}}{F_{u,in}}$ (acoustic counter part of $v_{BE}$) and the force output $\frac{F_{u,out}}{F_{u,in}}$ (acoustic counter part of $i_{CE}$). That is, we can control the wave output by choosing the operating point on this characteristic curve in a way similar to the electrical transistor. However, the main difference of the characteristic curve of the AT compared with the electrical one is that the control must have $\frac{F_{\theta,in}}{F_{u,in}}=({\frac{\widetilde{\Theta}}{\widetilde{U}}})_{L}$ to turn the AT off completely (again, it transmits the wave with zero control input as indicated with grey dotted line in Fig. \ref{fig_AT}a). 

Lastly, we show that the characteristic curve can be altered by employing different sets of design parameters. While there exists a variety of choices in terms of the design parameters as discussed in the previous section, we present a simple example in Fig. \ref{fig_AT}b, when we vary $\epsilon_u = 1, 2,$ and 3. When we have $\rho=0.5$ and $\epsilon_{\theta} = 0.078$, the turn off point moves towards left in the \textit{x}-axis as we introduce a larger $\epsilon_u$ (see the inset of Fig. \ref{fig_AT}b). The slope of the curves also changes as we adopt different $\epsilon_u$. This is plausible, since $\epsilon_u$ determines how much the dynamics in the control channel affects the wave transmission in the primary channel. Thus, we observe a steeper curve as we introduce a higher $\epsilon_u$. 

In this section, we numerically demonstrated that the PC system investigated in this study is capable of manipulating the primary channel's transmission efficiency through the control wave, which corresponds to the \textit{switching} function of conventional transistors. Another key feature of such conventional transistors is an amplification function, which enables manipulating a large-amplitude output signal using a small-amplitude control signal. The system proposed in this study meets this requirement in certain regimes. For instance, the slanted lines in Figs. \ref{fig_AT}a and \ref{fig_AT}b show the lines that denote the same magnitude between $F_{\theta, in}$ and $F_{u, out}$. We observe that the system in Fig. \ref{fig_AT}a does not meet this amplification requirement, while the ones in Fig. \ref{fig_AT}b successfully produce output waves' amplitude larger than that of the control wave. Thus, we numerically verify that the PC system equipped with a variable stiffness mechanism can be potentially exploited to be used as a mechanical counterpart of transistor systems.




 \section{Conclusions and Future Challenges}

In this study we proposed a coupled phononic crystal (PC) with a variable stiffness mechanism and investigated its wave dynamics via analytical and numerical approaches. The focus was placed on the mixing effect of mechanical waves propagating along primary and control channels. We verified that this wave mixing effect enables not only the control over the transmission efficiency of propagating waves, but also selective transmission of specific wave modes and eventually wave modulation effect. To the further extent, we showed the feasibility of realizing acoustic transistors, which fulfill both switching and amplification functionalities, using the proposed phononic system. 

This article proposed a number of possibilities within the study of PCs and can serve as a guide for several other avenues for future study.
Some of these potential future directions include the role of nonlinearity and the study of 
the additional structures that are made possible due to nonlinearity, such as traveling wave and breather solutions~\cite{khomeriki}.
Importantly, it should be mentioned that relevant mechanisms
may be considerably richer in higher dimensional settings, 
where the corresponding dispersion relation may enable
intriguing possibilities (including a dependence on the direction
or angle of propagation). 
Moreover, the interplay of nonlinearity with dimensionality
may be of interest, as the existence properties
of nonlinear states relevant to the present setting may be drastically 
different in higher dimensions~\cite{flachkl}.
Additionally, the extension to non-homogeneous systems would make a natural future direction.
Finally, the experimental verification of the system remains an open problem, but that is within reach given the state-of-art of PCs, which includes
 granular crystals \cite{Nester2001}, origami \cite{Yasuda}, and truss structures \cite{Amendola}. 
Such studies will be reported on in future publications.

 \section{Acknowledgements}

J.Y. and G.L. are grateful for the support from the ONR (N000141410388), NSF (CAREER-1553202) and Korea's ADD (UD140059JD). J.Y. and P.G.K. also acknowledge the support from the ARO (W911NF-15-1-0604). P.G.K. acknowledges support
also from the ERC under FP7, Marie Curie Actions, People, International
Research Staff Exchange (IRSES-605096).

\section*{Appendix}

\subsection*{Examples of coupled PCs with a variable stiffness mechanism}


\begin{figure}[htbp]
	\begin{center}
		\includegraphics[width=1\textwidth]{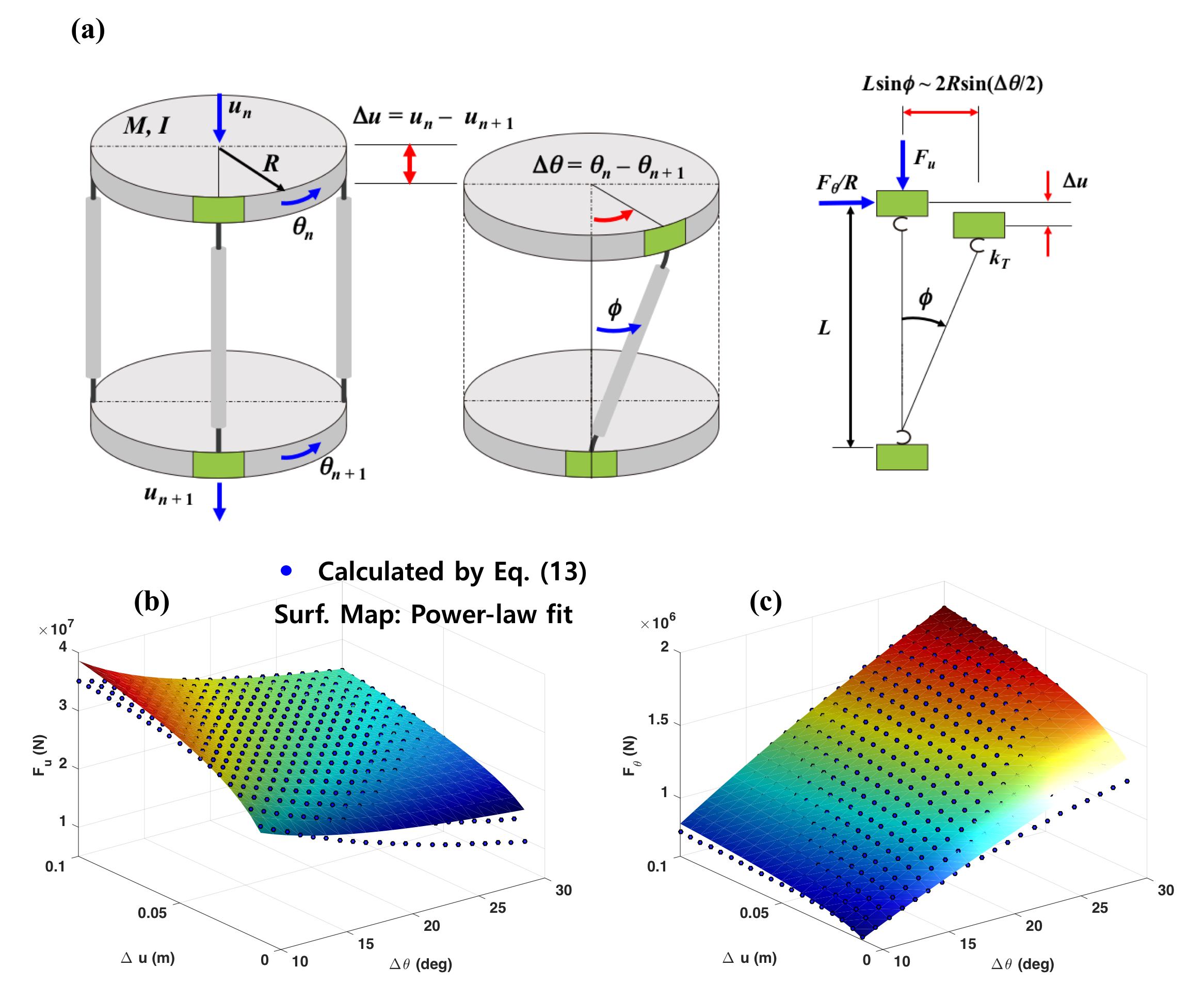}
		\caption{A prototypical system of the variable stiffness mechanism composed of buckled beam elements with (a) the schematic representation of the unit cell and its geometrical configurations. (b) The axial force $F_u$ with respect to the axial displacements $\Delta u = u_{n}-u_{n+1}$ and $\Delta\theta=\theta_{n}-\theta_{n+1}$; and (c) the torsional force $F_\theta$ with respect to the axial displacements $\Delta u = u_{n}-u_{n+1}$ and $\Delta\theta=\theta_{n}-\theta_{n+1}$ are plotted in blue dot based on Eq.~\eqref{eq:force_relation3}. The fitted results with power-law approximations are also plotted in surface maps.}
		\label{fig_beam}
	\end{center}
\end{figure}

Lumped masses connected with buckled beam elements can exhibit a variable stiffness mechanism (see Fig.~\ref{fig_beam}a). Here we consider a beam whose middle portion is rigid, while both ends can be buckled~\cite{Timoshenko} (Fig.~\ref{fig_beam}a). If we simplify the system as shown in the third image of Fig.~\ref{fig_beam}a assuming the buckled beam has a constant torsional stiffness $k_T$, we are able to write the potential energy $\psi$ as
\begin{equation}
	\label{eq:potential}
		\psi=\frac{1}{2} k_T \phi^2 - F_u \Delta u- F_\theta \Delta \theta,
\end{equation}
\noindent where $\phi$ represents the rotation angle of the beam element, and $F_u$ and $F_\theta$ are the axial and torsional forces, respectively. In Eq.~\eqref{eq:potential}, the first term represents the total elastic energy and the second and third terms denote the work done by external forces $F_u$ and $ F_\theta$. Under small perturbations, the geometrical relations among $\phi$, $\Delta u$, and $\Delta \theta$ are $\sin\phi \approx \frac{2R}{L}\sin{\frac{\Delta\theta}{2}}$ and $\cos\phi \approx 1-\frac{\Delta u}{L}$ (see Fig~\ref{fig_beam}a).
Thus, $\Delta\theta=2\sin^{-1}\left(\frac{1}{2R}\sqrt{\Delta u-(2L-\Delta u)}\right)$, 
where $L$ and $R$ are the length of the beam element and the radius of the lumped mass, respectively. 

By using the principle of minimum potential energy (i.e., $\frac{\partial\psi}{\partial \Delta u}=\frac{\partial\psi}{\partial \Delta u} \frac{\partial \Delta u}{\partial \Delta \theta}=0$ and $\frac{\partial\psi}{\partial \Delta \theta}=\frac{\partial\psi}{\partial \Delta \theta} \frac{\partial \Delta \theta}{\partial \Delta u}=0$), we are able to write $F_u$ and $F_\theta$ as function of $\Delta u$ and $\Delta \theta$:
\begin{eqnarray}
	\label{eq:force_relation1}
		F_u+A(\Delta u)F_\theta=B(\Delta u)k_T, \nonumber\\
		C(\Delta \theta)F_u+F_\theta=D(\Delta \theta)k_T.
\end{eqnarray}
\noindent Here $A(\Delta u)$, $B(\Delta u)$, $C(\Delta \theta)$, and $D(\Delta \theta)$ are expressed as
\begin{eqnarray}
	\label{eq:force_relation2}
		A(\Delta u)= & \frac{1}{R} \frac{L-\Delta u} { \sqrt { \Delta u (2L-\Delta u) \left[ 1 - \frac{\Delta u}{4R^2} (2L-\Delta u) \right] }}, \nonumber\\
		B(\Delta u)= & \frac{1}{L} \frac{\cos^{-1} \left( 1-\frac{\Delta u} {L} \right)} {\sqrt { \Delta u (2L-\Delta u) } }, \nonumber\\
		C(\Delta \theta)= & -\frac{1}{L} \frac{2R^2\sin \frac{\Delta \theta}{2} \cos\frac{\Delta \theta}{2}} {\sqrt {1-  \frac{4R^2}{L^2} \sin^2 \frac{\Delta \theta}{2} }}, \nonumber\\
		D(\Delta \theta)= & \frac{R}{L} \cos \frac{\Delta \theta}{2} \sin^{-1} \left( \frac{2R}{L} \sin \frac{\Delta \theta}{2} \right).
\end{eqnarray}
\noindent Solving Eq.~\eqref{eq:force_relation1} for $\Delta u$ and $\Delta \theta$, we obtain
\begin{eqnarray}
	\label{eq:force_relation3}
		F_u(\Delta u, \Delta \theta) = & \frac{1}{1-A(\Delta u)C(\Delta \theta)} \left[ B(\Delta u) -  A(\Delta u) D(\Delta \theta) \right] k_T, \nonumber\\
		F_\theta(\Delta u, \Delta \theta) = & \frac{1}{1-A(\Delta u)C(\Delta \theta)} \left[ -B(\Delta u) C(\Delta \theta) + D(\Delta \theta) \right] k_T.
\end{eqnarray}
\noindent Thus, the equation of motion can be expressed as
\begin{eqnarray}
	\label{eq:eom_beam}
		M\ddot{u}_{n} =& F_u(u_{n-1}-u_{n}, \theta_{n-1}-\theta_{n}) - F_u(u_{n}-u_{n+1}, \theta_{n}-\theta_{n+1}), \nonumber\\
		I\ddot{\theta}_{n} =& F_\theta(u_{n-1}-u_{n}, \theta_{n-1}-\theta_{n}) - F_\theta(u_{n}-u_{n+1}, \theta_{n}-\theta_{n+1}).
\end{eqnarray}

Figures~\ref{fig_beam}b-c plot the $F_u$ and $F_\theta$ as function of $\Delta u$ and $\Delta \theta$ (see blue dots). We also plot those relations with power-law fit, $K_a(\Delta u)^p(\Delta \theta)^r$ ($K_a=1.591 \times 10^{8}$, $p=0.216$, and $r=-0.399$) and $K_\theta(\Delta u)^q(\Delta \theta)^s$ ($K_\theta=2.091 \times 10^{5}$, $q=0.131$, and $r=0.726$) as surface maps in Fig.~\ref{fig_beam}b-c. Here we used $k_T = 1 \times 10^{7}$ N/deg, $L = 0.5$ m, and $ R = 0.2$ m. As we described in Section 2, these fitted data demonstrate that this post-buckled system can exhibit the variable stiffness mechanism in the form of a power-law. 

\begin{figure}[htbp]
	\begin{center}
		\includegraphics[width=1\textwidth]{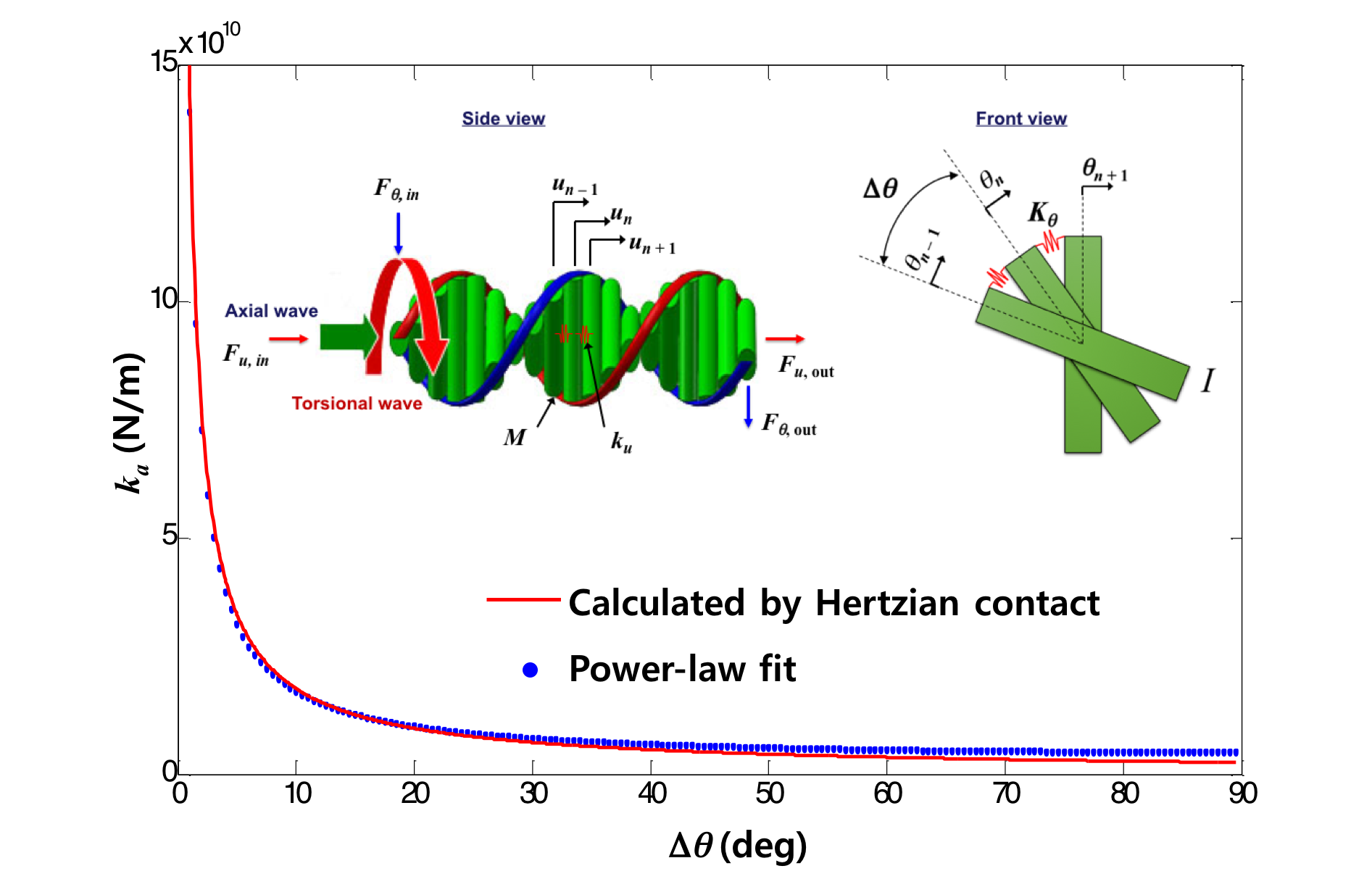}
		\caption{A prototypical system of the variable stiffness mechanism composed of helicoidally arranged cylindrical particles, which are in contact with each other via the Hertzian contact law that supports both axial and torsional waves. The axial stiffness $k_{a, (\theta_{n-1}-\theta_{n})}$ with respect to the torsional displacements $\Delta\theta=\theta_{n-1}-\theta_{n}$ is plotted in a red solid curve based on \eqref{eq:HPC_stiffness}. The fitted curve with power-law approximations is also plotted in blue dots. Insets show the geometrical configurations of the helicoidal phononic crystal in side and front views.} 
		\label{fig_HPC}
	\end{center}
\end{figure}

Another system that can exhibit the variable stiffness mechanism is a helicoidal phononic crystal (HPC), where the contact stiffness in the axial direction is given as a function of the relative angle between neighboring cells as below~\cite{HPC}:
\begin{equation}
	\label{eq:HPC_stiffness}
		k_{cyl}=\frac{\pi \sqrt{2} Y}{3 \left( 1-\nu^2 \right) e K(e)^{3/2}} \left \{ \left[ \left( \frac{1}{1-e^2} \right) E(e) - K(e) \right] \left[ K(e) - E(e) \right] \right \}^{1/4}
\end{equation}

Here, $Y$ represents the Young's modulus of HPC and $\nu$ the Poisson ratio. $K(e)$ and $E(e)$ are the complete elliptic integrals of the first and second kinds, respectively, and $e=\sqrt{ 1 - [\cos{\Delta \theta}/(1 + \cos{\Delta \theta})] ^{4/3} }$ is the eccentricity of the elliptical contact area between two cylindrical particles. By using $Y=72$ GPa and $\nu=0.17$, we find that the elliptic integral forms can be fit with the power law approximation, $k_{a, (\theta_{n-1}-\theta_{n})}=K_a(\theta_{n-1}-\theta_{n})^r=1.04 \times 10^8(\theta_{n-1}-\theta_{n})^{-0.55}$ as plotted in blue dots in Fig. \ref{fig_HPC}. When we consider the HPC connected with  linear torsional springs as depicted in the second inset in Fig. \ref{fig_HPC}, we can simply describe the torsional stiffness, $k_{\theta, (u_{n-1}-u_{n})}=K_\theta$. In this case, $p, q, r$ and $s$ in Eq. \eqref{eq:power} can be determined to be $p=1.5, q=0, r=-0.55$ and $s=1$ using the parameters used in this example. Note that we have zero $q$, since in this HPC, torsional waves affect the axial waves, but not vice versa.

\subsection*{Controllability of output wavelength via control frequency}

\begin{figure}[htbp]
	\begin{center}
		\includegraphics[width=1\textwidth]{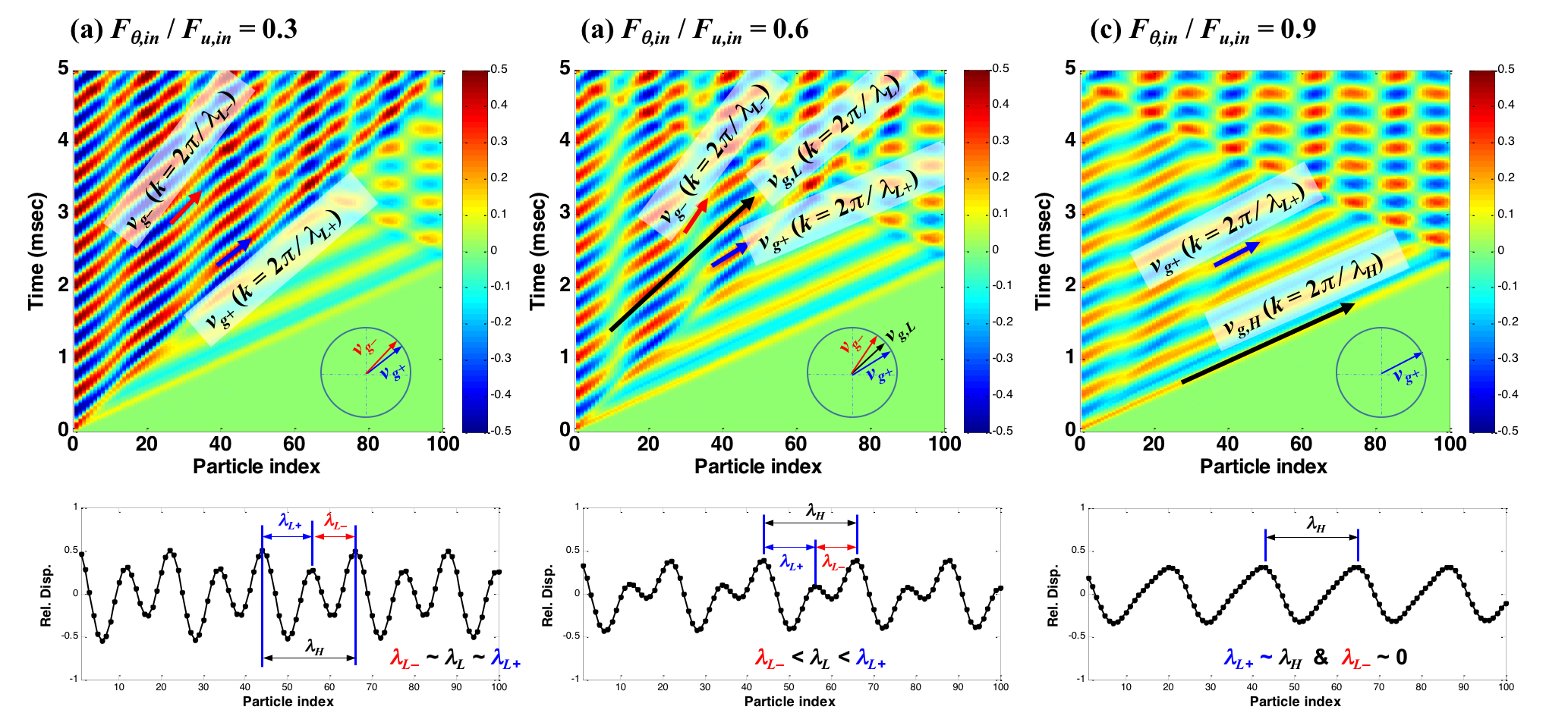}
		\caption{The evolution of wave profile in the space and time domain (color intensity represents the relative displacements) in the primary PC, showing the modulation effects under three different control inputs: (a) $\frac{F_{\theta,in}}{F_{u,in}}=0.3$, (b) $\frac{F_{\theta,in}}{F_{u,in}}=0.6$, and (c) $\frac{F_{\theta,in}}{F_{u,in}}=0.9$. The bottom panels of each color map capture the spatial wave profile at $t=40$ ms.}
		\label{fig_modulation}
	\end{center}
\end{figure}

As we discussed in Section 3.3, the modulation effect of the coupled PC generates two distinct wavelengths, $\lambda_{L-}$ and $\lambda_{L+}$, depending on the control input, $\frac{F_{\theta,in}}{F_{u,in}}$. To investigate how those wavelengths change with respect to the control input, we compare the harmonic responses of the system at $\Omega_{e,L}$ under three different control inputs; $\frac{F_{\theta,in}}{F_{u,in}}=0.3, 0.6,$ and 0.9. We plot the spatio-temporal maps of the particles' displacements in Fig. \ref{fig_modulation}a-\ref{fig_modulation}c. The bottom panels in Fig. \ref{fig_modulation} show the corresponding spatial wave profile at $t = 40$ ms. When $\frac{F_{\theta,in}}{F_{u,in}}=0.3$ (Fig.~\ref{fig_modulation}a), the two wavelengths are very close to $\lambda_{L}$ (the difference between two wavelengths, $\lambda_{L-}$ and $\lambda_{L+}$ is small). However, when inspecting the wave propagation in time and space domain, we can distinguish two different group velocities denoted as $v_{g-}$ (red arrow in Fig. \ref{fig_modulation}a) and $v_{g+}$ (blue arrow). The slopes of these arrows are calculated from the analytic dispersion in Eq. \eqref{eq:dispersion} (i.e., $\frac{\partial \Omega(k)}{\partial k}$) using the wavelengths $\lambda_{L-}$ and $\lambda_{L+}$ extracted from the bottom panel of Fig. \ref{fig_modulation}. As we increase the control input to $\frac{F_{\theta,in}}{F_{u,in}}=0.6$, the difference between the two wavelengths increases, and the corresponding group velocities become larger, as depicted in Fig. \ref{fig_modulation}b. Further increase of the control input makes the wavelength $\lambda_{L-}$ approach zero and $\lambda_{L+}$ approach $\lambda_H$, as shown in Fig. \ref{fig_modulation}c. In brief, our simulation reveals that as we increase the control input from the lower mode solution $\frac{F_{\theta,in}}{F_{u,in}}={\frac{\widetilde{\Theta}}{\widetilde{U}}}_{L}$ to the higher mode solution $\frac{F_{\theta,in}}{F_{u,in}}={\frac{\widetilde{\Theta}}{\widetilde{U}}}_{H}$, the shorter wavelength $\lambda_{L-}$ decreases from $\lambda_{L}$ to zero, while the longer wavelength $\lambda_{L+}$ increases from $\lambda_{L}$ to $\lambda_{H}$. This implies that we can generate different forms of modulated signals by simple manipulation of the control input. 

\subsection*{Existence of non-trivial cutoff frequencies}

\begin{figure}[htbp]
	\begin{center}
		\includegraphics[width=1\textwidth]{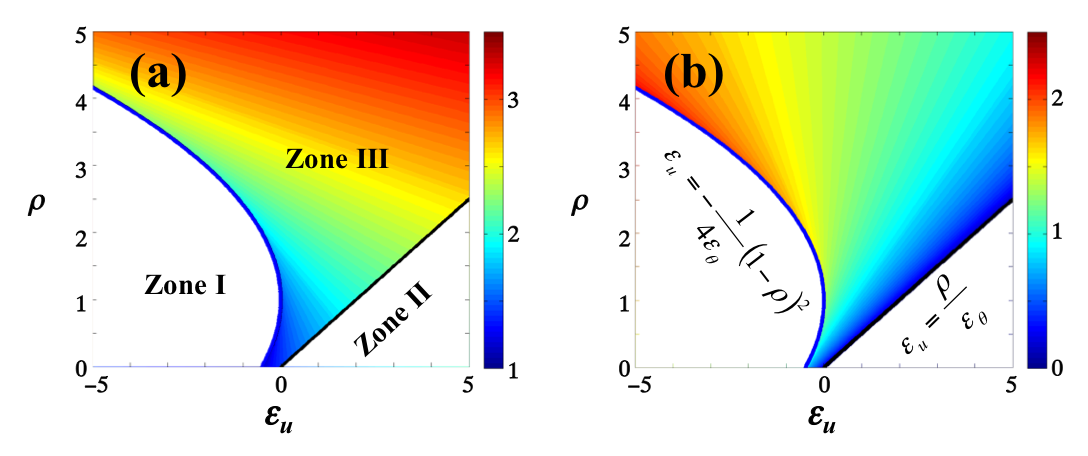}
		\caption{Contour plot of the two cut-off frequencies, (a) ${\Omega}_{C,H}$ and (b) ${\Omega}_{C,L}$, with respect to $\epsilon_u$ and $\rho$ for $\epsilon_{\theta}=0.5$. Three zones (I), (II), and (III) represent the status of zero, one, and two non-trivial cutoff frequencies.}
		\label{fig_cutoff_contour}
	\end{center}
\end{figure}

To investigate the design parameters of the coupled phononic crystal, we plotted surface maps of the system's frequencies as a function of design parameters in Section 4. To ease visualization, these maps can be plotted in a planar space, see Fig.~\ref{fig_cutoff_contour} for $\epsilon_{\theta}=0.5$. The higher and lower cutoff frequencies (i.e., ${\Omega}_{C,L}$ and ${\Omega}_{C,H}$) of the system are illustrated in Fig. \ref{fig_cutoff_contour}a and \ref{fig_cutoff_contour}b, respectively. We find that the highlighted zones are bounded by the aforementioned characteristic equations ($\epsilon_u = -\frac{1}{4\epsilon_{\theta}}(1-\rho)^2$ and $\epsilon_u = \frac{\rho}{\epsilon_{\theta}}$). We also see that depending on the combination of $\rho$ and $\epsilon_u$, we have three distinctive regions: (Zone I) non-existence of cutoff frequency, where the primary wave is not allowed to transmit; (Zone II) existence of a single cutoff frequency, where only the higher mode can be supported, and (Zone III) existence of two distinctive cutoff frequencies, where both higher and lower modes are supported (see Fig. \ref{fig_cutoff_contour}). Note that the particular case that we discussed in Section 3 (i.e., $\rho=1.5$, $\epsilon_u=1$ and $\epsilon_{\theta}=0.5$) belongs to the Zone III, where we can selectively choose and manipulate the efficiency of the two modes.


\section*{References}

\begin{thebibliography}{}


\bibitem{sound_atten} J. V. Sanchez-Perez, D. Caballero, R. Martinez-Sala, C. Rubio, J. Sanchez-Dehesa, F. Meseguer, J. Llinares, and F. Galvez, 1998, Sound Attenuation by a Two-Dimensional Array of Rigid Cylinders, Physical Review Letters, 80 (24), 5325--5328. doi:10.1103/PhysRevLett.80.5325

\bibitem{PC1} S.-C. S. Lin and T. J. Huang, 2011, Tunable phononic crystals with anisotropic inclusions, Physical Review B, 83 (17) 174303. doi:10.1103/PhysRevB.83.174303

\bibitem{PC4} R. K. Narisetti, M. Ruzzene, and M. J. Leamy, 2012, Study of wave propagation in strongly nonlinear periodic lattices using a harmonic balance approach, Wave Motion, 49 (2), 394-410. doi:10.1016/j.wavemoti.2011.12.005

\bibitem{PC2} J. Liu, Y. Wu, F. Li, P. Zhang, Y. Liu, and J. Wu, 2012, Anisotropy of homogenized phononic crystals with anisotropic material, Europhysics Letters, 98 (3), 36001. doi:10.1209/0295-5075/98/36001

\bibitem{mitigate1} G. Gantzounis, M. Serra-Garcia, K. Homma, J. M. Mendoza, and C. Daraio, 2013, Granular metamaterials for vibration mitigation, Journal of Applied Physics, 114 (9), 093514. doi:10.1063/1.4820521

\bibitem{PC3} S. Babaee, P. Wang, and K. Bertoldi, 2015, Three-dimensional adaptive soft phononic crystals, Journal of Applied Physics, 117 (24), 244903. doi:10.1063/1.4923032

\bibitem{gran1} S. Sen, J. Hong, J. Bang, E. Avalos, and R. Doney, 2008, Solitary waves in the granular chain, Physics Reports, 462 (2), 21-66. doi:10.1016/j.physrep.2007.10.007

\bibitem{gran2} M. A. Porter, C. Daraio, E. B. Herbold, I. Szelengowicz, and P. G. Kevrekidis, 2008, Highly nonlinear solitary waves in periodic dimer granular chains, Physical Review, E 77 (1), 015601. doi:10.1103/PhysRevE.77.015601

\bibitem{PC5} F. Casadei and M. Ruzzene, 2012, Frequency-domain bridging method for the analysis of wave propagation in damaged structures, Wave Motion, 49 (6), 605-616. doi:10.1016/j.wavemoti.2012.04.001

\bibitem{Martin} M. Maldovan, 2013, Sound and heat revolutions in phononics, Nature, 503 (7475), 209-217. doi:10.1038/nature12608

\bibitem{local_reson1} Z. Liu, X. Zhang, Y. Mao, Y. Y. Zhu, Z. Yang, C. T. Chan, and P. Sheng, 2000, Locally Resonant Sonic Materials, Science, 289 (5485), 1734-1736. doi:10.1126/science.289.5485.1734

\bibitem{switch1} F. Li, P. Anzel, J. Yang, P. G. Kevrekidis, and C. Daraio, 2014, Granular acoustic switches and logic elements, Nature Communications, 5, 5311. doi:10.1038/ncomms6311

\bibitem{local_reson2} F. Casadei and K. Bertoldi, 2014, Wave propagation in beams with periodic arrays of airfoil-shaped resonating units, Journal of Sound and Vibration, 333 (24), 6532-6547. doi:10.1016/j.jsv.2014.07.008

\bibitem{local_reson3} P. Wang, F. Casadei, S. H. Kang, and K. Bertoldi, 2015, Locally resonant band gaps in periodic beam lattices by tuning connectivity, Physical Review B, 91 (2), 020103. doi:10.1103/PhysRevB.91.020103

\bibitem{switch2} N. Boechler, G. Theocharis, and C. Daraio, 2011, Bifurcation-based acoustic switching and rectification, Nature Materials, 10 (9), 665-668. doi:10.1038/nmat3072

\bibitem{TET1} M. A. Hasan, Y. Starosvetsky, A. F. Vakakis, and L. I. Manevitch, 2013, Nonlinear targeted energy transfer and macroscopic analog of the quantum Landau?Zener effect in coupled granular chains, Physica D: Nonlinear Phenomena, 252, 46-58. doi:10.1016/j.physd.2013.02.011

\bibitem{2D_PC} H. Pichard, A. Duclos, J. P. Groby, V. Tournat, and V. E. Gusev, 2012, Two-dimensional discrete granular phononic crystal for shear wave control, Physical Review B, 86 (13), 134307. doi:10.1103/PhysRevB.86.134307

\bibitem{HPC} F. Li, C. Chong, J. Yang, P. G. Kevrekidis, and C. Daraio, 2014, Wave transmission in time- and space-variant helicoidal phononic crystals, Physical Review E, 90 (5), 053201. doi:10.1103/PhysRevE.90.053201

\bibitem{3D_wood} E. Kim, Y. H. N. Kim, and J. Yang, 2015, Nonlinear stress wave propagation in 3D woodpile elastic metamaterials, International Journal of Solids and Structures, 58, 128-135. doi:http://dx.doi.org/10.1016/j.ijsolstr.2014.12.024

\bibitem{TET2} Y. Zhang, M. Hasan, Y. Starosvetsky, and A. Vakakis, 2015, Nonlinear mixed solitary?Shear waves and pulse equi-partition in a granular network, Physica D: Nonlinear Phenomena, 291, 45-61. doi:10.1016/j.physd.2014.10.005

\bibitem{brightb} N. Boechler, G. Theocharis, S. Job, P. G. Kevrekidis, Mason A. Porter, and C. Daraio, 2010, Discrete breathers in one-dimensional diatomic granular crystals, Physical Review Letters, 104 (24), 244302. doi:10.1103/PhysRevLett.104.244302

\bibitem{darkb} C. Chong, F. Li, J. Yang, M. O. Williams, I. G. Kevrekidis, P. G. Kevrekidis, and C. Daraio, 2014, Damped-driven granular chains: an ideal playground for dark breathers and multibreathers, Physical Review E, 89 (3), 032924. doi:10.1103/PhysRevE.89.032924

\bibitem{tune1} A. Khelif, P. A. Deymier, B. Djafari-Rouhani, J. O. Vasseur, and L. Dobrzynski, 2003, Two-dimensional phononic crystal with tunable narrow pass band: Application to a waveguide with selective frequency, Journal of Applied Physics, 94 (3), 1308-1311. doi:10.1063/1.1557776

\bibitem{tune2} M. Meidani, E. Kim, F. Li, J. Yang, and D. Ngo, 2015, Tunable evolutions of wave modes and bandgaps in quasi-1D cylindrical phononic crystals, Journal of Sound and Vibration, 334, 270-281. doi:http://dx.doi.org/10.1016/j.jsv.2014.09.010

\bibitem{tune3} A. Bergamini, T. Delpero, L. D. Simoni, L. D. Lillo, M. Ruzzene, and P. Ermanni, 2014, Phononic Crystal with Adaptive Connectivity, Advanced Materials, 26 (9), 1343-1347. doi:10.1002/adma.201305280

\bibitem{Psarobas_2014} I. E. Psarobas, D. A. Exarchos, and T. E. Matikas, 2014, Birefringent phononic structures. AIP Advances, 4 (12), 124307. doi:10.1063/1.4904812

\bibitem{Gonella_PRL2015} R. Ganesh, S. Gonella, 2015, From modal mixing to tunable functional switches in nonlinear phononic crystals, Physical Review Letters, 114 (5), 054302. doi:10.1103/PhysRevLett.114.054302

\bibitem{khomeriki} M. Malishava and R. Khomeriki, 2015, All-Phononic Digital Transistor on the Basis of Gap-Soliton Dynamics in an Anharmonic Oscillator Ladder, Physical Review Letters, 115 (10), 104301. doi:10.1103/PhysRevLett.115.104301

\bibitem{var_stiff1} Y.-S. Wu and C.-C. Lan, 2014, Linear Variable-Stiffness Mechanisms Based on Preloaded Curved Beams, Journal of Mechanical Design, 136 (12), 122302. doi:10.1115/1.4028705

\bibitem{var_stiff2} S. Nagarajaiah and S. Sahasrabudhe, 2006, Seismic response control of smart sliding isolated buildings using variable stiffness systems: an experimental and numerical study, Earthquake Engineering \& Structural Dynamics, 35 (2), 177-197. doi:10.1002/eqe.514

\bibitem{compliant1} G. Palli, G. Berselli, C. Melchiorri, and G. Vassura, 2011, Design of a Variable Stiffness Actuator Based on Flexures, Journal of Mechanisms and Robotics, 3 (3), 034501. doi:10.1115/1.4004228

\bibitem{compliant2} C.-C. Lan and Y.-J. Cheng, 2008, Distributed Shape Optimization of Compliant Mechanisms Using Intrinsic Functions, Journal of Mechanical Design, 130 (7), 072304. doi:10.1115/1.2890117

\bibitem{compliant3} R. Ham, T. G. Sugar, B. Vanderborght, K. W. Hollander, and D. Lefeber, 2009, Compliant actuator designs, Robotics \& Automation Magazine, IEEE, 16 (3), 81-94. doi:10.1109/MRA.2009.933629

\bibitem{origami} X. Zhao, Y. Hu, and I. Hagiwara, 2011, Shape Optimization to Improve Energy Absorption Ability of Cylindrical Thin-Walled Origami Structure, Journal of Computational Science and Technology, 5 (3), 148-162. doi:10.1299/jcst.5.148

\bibitem{Timoshenko} J. D. Aristizabal-Ochoa, 2007, Large deflection and postbuckling behavior of Timoshenko beam columns with semi-rigid connections including shear and axial effects, Engineering Structures, 29 (6), 991-1003. doi:10.1016/j.engstruct.2006.07.012

\bibitem{Timoshenko_book} S, P, Timoshenko, \emph{Theory of Elastic Stability} (McGraw-Hill, New York, 1961).

\bibitem{Landau} L. D. Landau and E. M. Lifshitz, \emph{Theory of Elasticity} (Pergamon, NY, 1970).

\bibitem{Hertzian2} K. L. Johnson,\emph{Contact Mechanics} (Cambridge University Press, Cambridge, 1985).

\bibitem{TFT1} K.E. Atkinson, An Introduction to Numerical Analysis, 2nd edn. (John Wiley and Sons, New York, 1989).

\bibitem{BJT} Robert F. Pierret, \emph{Semiconductor Device Fundamentals} (Addison Wesley, 2nd edition, Addison Wesley, 1996).

\bibitem{flachkl} S. Flach, K. Kladko, and R. S. MacKay,1997, Energy Thresholds for Discrete Breathers in One-, Two-, and Three-Dimensional Lattices, Physical Review Letters, 78 (7), 1207-1210.  doi:10.1103/PhysRevLett.78.1207

\bibitem{Nester2001}V. F. Nesterenko, \emph{Dynamics of Heterogeneous Materials} (Springer-Verlag, New York, 2001).

\bibitem{Yasuda} H. Yasuda and J. Yang, 2015, Reentrant Origami-Based Metamaterials with Negative Poisson's Ratio and Bistability, Physical Review Letters, 114 (18), 185502. doi:10.1103/PhysRevLett.114.185502

\bibitem{Amendola} F. Fraternali, G. Carpentieri, A. Amendola, R.E. Skelton and V.F. Nesterenko, 2014, Multiscale tunability of solitary wave dynamics in tensegrity metamaterials, Applied Physics Letters, 105 (20), 201903. doi:10.1063/1.4902071




\end{thebibliography}
\end{document}